\documentclass{book}

\usepackage[paperwidth=17cm, paperheight=24cm, includeheadfoot,
            left=2cm, right=2cm, top=2cm, bottom=1.3cm, twoside]{geometry}

\usepackage%
 {
   algorithm2e, amsmath, amssymb, appendix, caption, cite, enumitem, fancyhdr,
  fancyvrb, graphicx, import, lineno, longtable, remreset, shorttoc, sidecap,
  url
 }

\newcommand{\tmpstring}{}
\newcommand{\settmpstring}[1]{\renewcommand{\tmpstring}{#1}}
\newcommand{\SourceCodeLines}[1]
 {%
  \settmpstring{{\ttfamily\bfseries\tiny\theFancyVerbLine}}
  \ifnum#1>9
    \settmpstring
     {\parbox[b]{7.5pt}{\ttfamily\bfseries\tiny\rightline\theFancyVerbLine}}
  \fi
  \ifnum#1>99
    \settmpstring
     {\parbox[b]{11.2pt}{\ttfamily\bfseries\tiny\rightline\theFancyVerbLine}}
  \fi
 }

\def\thepart{\Alph{part}}

\makeatletter
\@removefromreset{figure}{chapter}
\makeatother

\makeatletter
\renewcommand{\thefigure}{\@arabic\c@figure}
\makeatother

\makeatletter
\@removefromreset{table}{chapter}
\makeatother

\makeatletter
\renewcommand{\thetable}{\@arabic\c@table}
\makeatother

\makeatletter
\@removefromreset{equation}{chapter}
\makeatother

\makeatletter
\renewcommand{\theequation}{\@arabic\c@equation}
\makeatother

\usepackage{tocloft}
\usepackage{minitoc}

\cftsetpnumwidth{6.1mm}

\makeatletter
\renewcommand{\@tocrmarg}{4em}
\makeatother

\mtcsetformat{minitoc}{tocrightmargin}{2.55em plus 1fil}

\newcommand{\Author}{}
\newcommand{\AuthorLastName}[1]{\renewcommand{\Author}{#1}}

\def\Title#1{\chapter[\thepart\thelecture\ \protect\mbox{\protect%
\parbox[t]{110mm}{#1 \textit{(\Author)}}}\smallskip]{\Large #1}}

\newsavebox{\shortTitleBox}
\def\shortTitle#1{\savebox{\shortTitleBox}{#1 \textit{(\Author)}}}

\newcounter{lecture}[part]

\fancypagestyle{plain}
 {
  \renewcommand{\headrulewidth}{0pt}
  \fancyhead[OL]
   {
    \vspace{-8.5pt}
    \textit{Lecture given at the International Summer School
   \emph{Modern Computational Science}}\\
    \textit{(August 9-20, 2010, Oldenburg, Germany)}
  }
 }

\fancypagestyle{plain}
 {
  \renewcommand{\headrulewidth}{0pt} 
  \fancyhf{}
 }

\fancypagestyle{empty}
 {
  \renewcommand{\headrulewidth}{0pt}
  \fancyhead{}
 }

 {
  \fancyhf{}
  \fancyhead[OR]{\rightmark}
 }

\newcommand{\SwitchToFancy}
 {%
  \pagestyle{fancy}
   {
    \renewcommand{\headrulewidth}{0.4pt}
    \fancyhf{}
    \fancyhead[OR]{\rightmark}
    \fancyfoot[OR]{\thepage}
    \fancyfoot[EL]{\thepage}
    \fancyhead[EL]{\usebox{\shortTitleBox}}
   }
 }

\makeatletter
\def\thebibliography#1
{
 \section*{References\@mkboth{References}{References}}
 \list{[\arabic{enumi}]}
  {
   \settowidth\labelwidth{[#1]}\leftmargin\labelwidth\advance\leftmargin
   \labelsep\usecounter{enumi}
  }
 \def\newblock{\hskip .11em plus .33em minus .07em}
 \sloppy\clubpenalty4000\widowpenalty4000\sfcode`\.=1000\relax
}
\makeatother

\makeatletter
\renewcommand{\@makefntext}[1]{\setlength{\parindent}{0pt}%
\begin{list}{}{\setlength{\labelwidth}{1.5em}%
\setlength{\leftmargin}{\labelwidth}%
\setlength{\labelsep}{3pt}\setlength{\itemsep}{0pt }%
\setlength{\parsep}{0pt}\setlength{\topsep}{0pt}%
\footnotesize}\item[\hfill\@makefnmark]#1%
\end{list}}
\makeatother

\begin{document}

\SwitchToFancy

\dominitoc

\faketableofcontents

\renewcommand{\cftsecfont}{\bfseries}
\renewcommand{\cftsecleader}{\bfseries\cftdotfill{\cftdotsep}}
\renewcommand{\cftsecpagefont}{\bfseries}

\setlength{\cftsubsecindent}{12.5mm}

\captionsetup{width=0.9\textwidth,font=small,labelfont=bf}


\stepcounter{lecture}
\setcounter{figure}{0}
\setcounter{equation}{0}
\setcounter{table}{0}


\AuthorLastName{Katzgraber}

\Title{Introduction to Monte Carlo Methods}

\shortTitle{Monte Carlo Methods}

\SwitchToFancy

\bigskip
\bigskip


\begin{raggedright}
  \itshape Helmut G.~Katzgraber\\
  \bigskip
  Department of Physics and Astronomy, Texas A\&M University\\
  College Station, Texas 77843-4242 USA\\
  \medskip
  Theoretische Physik, ETH Zurich\\
  CH-8093 Zurich, Switzerland\\
  \bigskip
  \bigskip
\end{raggedright}


\paragraph{Abstract.} Monte Carlo methods play an important role in
scientific computation, especially when problems have a vast phase
space. In this lecture an introduction to the Monte Carlo method is
given. Concepts such as Markov chains, detailed balance, critical
slowing down, and ergodicity, as well as the Metropolis algorithm
are explained. The Monte Carlo method is illustrated by numerically
studying the critical behavior of the two-dimensional Ising ferromagnet
using finite-size scaling methods. In addition, advanced Monte Carlo
methods are described (e.g., the Wolff cluster algorithm and parallel
tempering Monte Carlo) and illustrated with nontrivial models from
the physics of glassy systems.  Finally, we outline an approach to
study rare events using a Monte Carlo sampling with a guiding function.


\minitoc

\section{Introduction}
\label{sec:introMC}

The Monte Carlo method in computational physics is possibly one of the
most important numerical approaches to study problems spanning {\em
all} thinkable scientific disciplines. The idea is seemingly simple:
Randomly {\em sample} a volume in $d$-dimensional space to obtain an
estimate of an {\em integral} at the price of a statistical error. For
problems where the phase space dimension is very large---this is
especially the case when the dimension of phase space depends on the
number of degrees of freedom---the Monte Carlo method outperforms
any other integration scheme. The difficulty lies in smartly choosing
the random samples to minimize the numerical effort.

The term {\em Monte Carlo} method was coined in the 1940s by physicists
S.~Ulam, E.~Fermi, J.~von Neumann, and N.~Metropolis (amongst others)
working on the nuclear weapons project at Los Alamos National Laboratory.
Because random numbers (similar to processes occurring in a casino,
such as the Monte Carlo Casino in Monaco) are needed, it is believed
that this is the source of the name.  Monte Carlo methods were central
to the simulations done at the Manhattan Project, yet mostly hampered
by the slow computers of that era. This also spurred the development of
fast random number generators, discussed in another lecture of this series.

In this lecture, focus is placed on the standard Metropolis algorithm
to study problems in statistical physics, as well as a variation
known as exchange or parallel tempering Monte Carlo that is very
efficient when studying problems in statistical physics with complex
energy landscapes (e.g., spin glasses, proteins, neural networks)
\cite{comment:sources}. In general, continuous phase transitions are
discussed. First-order phase transitions are, however, beyond the
scope of these notes.

\section{Monte Carlo integration}
\label{sec:integration}

The motivation for Monte Carlo integration lies in the fact that most
standard integration schemes fail for high-dimensional integrals. At
the same time, the space dimension of the phase space of typical
physical systems is very large. For example, the phase space dimension
for $N$ classical particles in three space dimensions is $d = 6N$
(three coordinates and three momentum components are needed to fully
characterize a particle). This is even worse for the case of $N$
classical Ising spins (discussed below) which can take the values
$\pm 1$. In this case the phase space dimension is $2^N$, a number
that grows exponentially fast with the number of spins!  Therefore,
integration schemes such as Monte Carlo methods, where the error is
independent of the space dimension, are needed.

\subsection{Traditional integration schemes}
\label{sec:traditional}

\begin{figure}[!htb]
\centering
\includegraphics[scale=0.50]{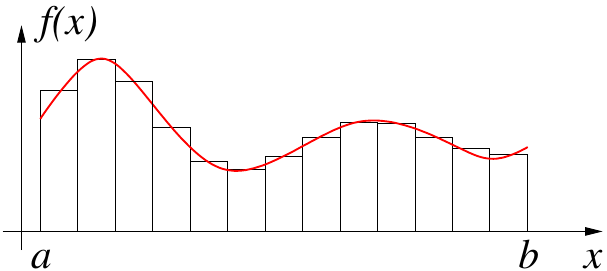}
\caption
{
Illustration of the midpoint rule. The integration interval $[a,b]$ is
divided into $M$ slices, the area of each slice approximated by the
width of the slice, $\delta = (b - a)/M$, times the function evaluated
at the midpoint of each slice.
}
\label{fig:midp}
\end{figure}

Before introducing Monte Carlo integration, let us review standard
integration\linebreak schemes to highlight the advantages of random sampling
methods. In general, the goal is to compute the following {\em
one-dimensional} integral
\begin{equation}
I = \int_a^b f(x) dx\;.
\label{eq:integral}
\end{equation}
Traditionally, one partitions the interval $[a,b]$ into $M$ slices
of width $\delta = (b - a)/M$ and then performs a $k$th order
interpolation of the function $f(x)$ for each interval to approximate
the integral as a discrete sum (see Fig.~\ref{fig:midp}). For example,
to first order, one performs the {\em midpoint rule} where the area
of the $l$th slice is approximated by a rectangle of width $\delta$
and height $f[(x_l + x_{l+1})/2]$. It follows that
\begin{equation}
I \approx \sum_{l = 0}^{M-1}  \delta \cdot f[(x_l + x_{l+1})/2]\;.
\label{eq:slice}
\end{equation}
For $M \to \infty$ the discrete sum converges to the integral of
$f(x)$.  Convergence can be improved by replacing the rectangle with a
linear interpolation between $x_l$ and $x_{l+1}$ (trapezoidal rule) or
a weighted quadratic interpolation (Simpson's rule) \cite{press:95}.
One can show that the error made due to the approximation of the
function is proportional to $\sim M^{-1}$ for the midpoint rule if the
function is evaluated at one of the interval's edges (in the center
as shown above $\sim M^{-2}$), $\sim M^{-2}$ for the trapezoidal rule,
and $\sim M^{-4}$ for Simpson's rule.  The convergence of the midpoint
rule can thus be slow and the method should be avoided.

A problem arises when a multi-dimensional integral needs to be
computed.  In this case one can show that, for example, the error
of Simpson's rule scales as $\sim M^{-4/d}$ because each space
component has to be partitioned independently.  Clearly, for space
dimensions larger than $4$ convergence becomes very slow. Similar
arguments apply for any other traditional integration scheme where the
error scales as $\sim M^{-\kappa}$: if applied to a $d$-dimensional
integral the error scales $\sim M^{-\kappa/d}$.

\subsection{Simple and Markov-chain sampling}

\begin{SCfigure}[1.2][!htb]
  \includegraphics[scale=0.70]{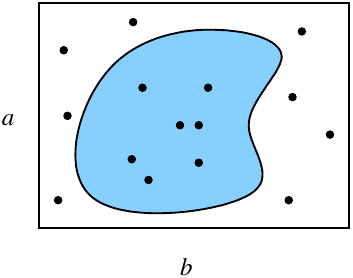}\hspace{2pc}
  \caption
   {
    Illustration of simple-sampling Monte Carlo integration. An
    unknown area (pond) is enclosed by a rectangle of known area
    $A = ab$. By randomly sampling the area with pebbles, a statistical
    estimate of the pond's area can be computed.
    \vspace{1.5pc}
   }
  \label{fig:pond}
\end{SCfigure}

One way to overcome the limitations imposed by high-dimensional
volumes is {\em simple sampling} Monte Carlo. A simple analogy is to
determine the area of a pond by throwing rocks. After enclosing the
pond with a known area (e.g., a rectangle) and having enough beer
or wine \cite{comment:alk}, pebbles are randomly thrown into the
enclosed area.  The ratio of stones in the pond and the total number
of thrown stones is a {\em simple sampling} statistical estimate for
the area of the pond, see Fig.~\ref{fig:pond}.

\begin{SCfigure}[1.2][!htb]
  \includegraphics[scale=0.70]{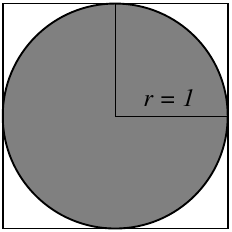}\vspace{1.5pc}\hspace{2pc}
  \caption
   {
    Monte Carlo estimate of $\pi$ by randomly sampling the unit circle:
    two random numbers $x$ and $y$ in the range $[0,1]$ are computed. If
    $x^2 + y^2 \le 1$, the resulting point is in the unit circle.  After $M$
    trials an estimate of $\pi/4$ can be computed with a statistical error
    $\sim M^{-1/2}$.
   }
  \label{fig:circle}
\end{SCfigure}

A slightly more ``scientific'' example is to compute $\pi$ by applying
Monte Carlo integration to the unit circle. The area of the unit circle
is given by $A_\circ = \pi r^2$ with $r = 1$; the top right quadrant
can be enclosed by a square of size $r$ and area $A_\Box = r^2$
(see Fig.~\ref{fig:circle}). An estimate of $\pi$ can be accomplished
with the following pseudo-code algorithm \cite{comment:pseudo}
that performs a simple sampling of the top-right quadrant:

\SourceCodeLines{99}
\begin{Verbatim}[fontsize=\small]
 algorithm simple_pi
     initialize n_hits     0
     initialize m_trials   10000
     initialize counter    0

     while(counter < m_trials) do
         x = rand(0,1)
         y = rand(0,1)
         if(x**2 + y**2 < 1) 
             n_hits++
         fi
         counter++
     done
 
 return pi = 4*n_hits/m_trials
\end{Verbatim}

\noindent For each of the \texttt{m\_trials} trials we generate
two uniform random numbers \cite{press:95} in the interval $[0,1]$
[with \texttt{rand(0,1)}] and test in line 9 of the algorithm if
these lie in the unit circle or not. The counter \texttt{n\_hits}
is then updated if the resulting number is {\em in} the circle. In
line 15 a statistical estimate of $\pi$ is then returned.

\begin{SCfigure}[][!htb]
  \includegraphics[scale=0.70]{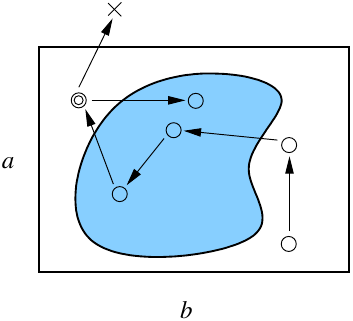}\hspace{2pc}
  \caption
   {
    Illustration of Markov-chain Monte Carlo. The new state is always
    derived from the previous state. At each step a pebble is thrown
    in a random direction, the following throw has its origin at the
    landing position of the previous one. If a pebble lands outside
    the rectangular area (cross) the move is rejected and the last
    position recorded twice (double circle).
   }
  \vspace{1pc}
  \label{fig:pondm}
\end{SCfigure}

Before applying these ideas to the integration of a function, we
introduce the concept of a {\em Markov chain} \cite{metropolis:49}. In
the simple-sampling approach to estimate the area of a pond as presented
above, the random pebbles used are independent in the sense that
a newly-selected pebble to be thrown into the rectangular area
in Fig.~\ref{fig:pond} does not depend in any way on the position
of the previous pebbles.  If, however, the pond is very large, it
is impossible to throw pebbles randomly from one position. Thus
the approach is modified: After enough beer you start at a random
location (make sure to drain the pond first) and throw a pebble into
a random direction.  You then walk to that pebble, pull a new pebble
out of a pebble bucket you have with you and repeat the operation.
This is illustrated in Fig.~\ref{fig:pondm}. If the pebble lands
{\em outside} the rectangular area, the thrower should go get the
outlier and place it on the {\em current} position of the thrower,
i.e., if the move lies outside the sampled area, it is {\em rejected}
and the last move counted twice. Why? This will be explained later
and is called {\em detailed balance} (see p.\ \pageref{page:detbal}).
Basically, it ensures that the Markov chain is reversible.
After many beers and throws, pebbles are
scattered around the rectangular area, with small piles of multiple
pebbles closer to the boundaries (due to rejected moves).

Again, these ideas can be used to estimate $\pi$ by Markov-chain
sampling the unit circle. Later, the Metropolis algorithm, which is
based on these simple ideas, is introduced in detail using models from
statistical physics. The following algorithm describes Markov-chain Monte
Carlo for estimating $\pi$:

\SourceCodeLines{99}
\begin{Verbatim}[fontsize=\small]
 algorithm markov_pi
     initialize n_hits     0
     initialize m_trials   10000
     initialize x          0
     initialize y          0
     initialize counter    0

     while(counter < m_trials) do
         dx = rand(-p,p)
         dy = rand(-p,p)
         if(|x + dx| < 1 and |y + dy| < 1) 
             x = x + dx
             y = y + dy
         fi
         if(x**2 + y**2 < 1) 
             n_hits++
         fi
         counter++
     done

 return pi = 4*n_hits/m_trials
\end{Verbatim}

\noindent The algorithm starts from a given position in the space to
be sampled [here $(0,0)$] and generates the position of the new dot
from the position of the previous one. If the new position is outside
the square, it is rejected (line 11). A careful selection of
the step size \texttt{p} used to generate random numbers in the range
$[-p,p]$ is of importance: When \texttt{p} is too small, convergence
is slow, whereas if \texttt{p} is too large many moves are rejected
because the simulation will often leave the unit square. Therefore,
a value of \texttt{p} has to be selected such that consecutive moves
are accepted approximately 50\% of the time. 

The simple-sampling approach has the advantage over the Markov-chain
approach in that the different samples are independent and thus
not correlated.  In the Markov-chain approach the new state depends
on the previous state. This can be a problem since there might be a
``memory'' associated with this behavior. If this memory is large,
then the {\em autocorrelation times} (i.e., the time it takes the
system to forget where it was) are large and many moves have to
be discarded. Then why even think about the Markov-chain approach?
Because in the study of physical systems it is generally easier to
slightly (and randomly) change an existing state than to generate a
new state from scratch for each step of the calculation. For example,
when studying a system of $N$ spins it is easier to flip one spin
according to a given probability distribution than to generate a
new configuration from scratch with a pre-determined probability
distribution.

Let us apply now these ideas to perform a simple-sampling estimate
of the integral of an actual function. As an example, we select a simple
function, namely
\begin{equation}
f(x) = x^n 
\;\;\;\;\;\;\;
\rightarrow
\;\;\;\;\;\;\;
I = \int_0^1 f(x) dx
\label{eq:function}
\end{equation}
with $n > -1$. Using simple-sampling Monte Carlo, the integral can be
estimated via

\SourceCodeLines{99}
\begin{Verbatim}[fontsize=\small]
 algorithm simple_integrate
     initialize integral   0
     initialize m_trials   10000
     initialize counter    0

     while(counter < m_trials) do
         x = rand(0,1)
         integral += x**n 
         counter++
     done
 
 return integral/m_trials
\end{Verbatim}

\noindent In line 8 we evaluate the function at the random location
and add the result to the estimate of the integral, i.e.,
\begin{equation}
I \approx \frac{1}{M}\sum_i^{M}f(x_i)\;,
\end{equation}
where we have set \texttt{m\_trials} $= M$. To calculate the error of
the estimate, we need to compute the variance of the function. For
this we need to also perform a simple sampling of the square of the
function, i.e., add a line to the code with \texttt{integral\_square
+= x**(2*n)}. It then follows \cite{krauth:06} for the statistical
error of the integral $\delta I$
\begin{equation}
\delta I = \sqrt{\frac{{\rm Var}f}{M-1}}\;,
\;\;\;\;\;\;
\;\;\;\;\;\;
\;\;\;\;\;\;
{\rm Var}f = \langle f^2 \rangle - \langle f \rangle^2,
\label{eq:variance}
\end{equation}
with 
\begin{equation}
\langle f^k \rangle = \int_0^1 [f(x)]^k dx \approx
\frac{1}{M}\sum_i^M [f(x_i)]^k\;.
\label{eq:moments}
\end{equation}
Here $x_i$ are uniformly distributed random numbers.  The important
detail is that Eq.~(\ref{eq:variance}) does {\em not} depend on
the space dimension and merely on $M^{-1/2}$. This means that, for
example, for space dimensions $d > 8$ Monte Carlo sampling outperforms
Simpson's rule.

\begin{SCfigure}[1.2][!htb]
  \includegraphics[scale=0.50]{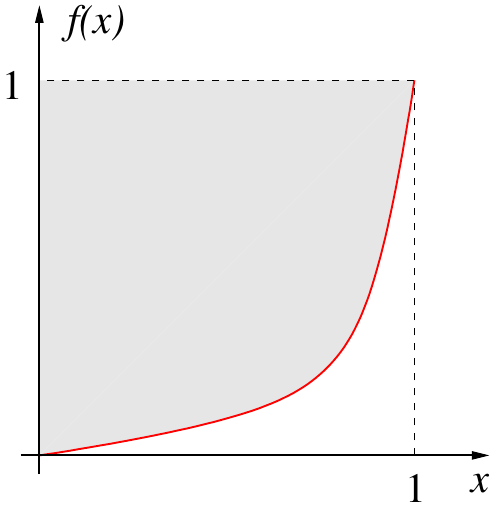}\vspace{1pc}\hspace{3pc}
  \caption
   {
    Illustration of the simple-sampling approach when integrating $f(x)
    = x^n$ with $n \gg 1$. The function has most support for $x \to1$.
    Because random numbers are generated with a uniform probability,
    the whole range $[0,1]$ is sampled equally probable, although for
    $x \to 0$ the contribution to the integral is small. Thus, the integral
    converges slowly.
    \vspace*{2pc}
   }
  \label{fig:reject}
\end{SCfigure}

The presented simple-sampling approach has one crucial problem: When
in the example shown the exponent $n$ is close to $-1$ or much larger
than $1$ the variance of the function in the interval is large.  At the
same time, the interval $[0,1]$ is sampled uniformly.  Therefore,
similar to the estimate of $\pi$, areas which carry little weight for
the integral are sampled with equal probability as areas which carry
most of the function's support (see Fig.~\ref{fig:reject}). Therefore
the integral and error converge slowly.  To alleviate the situation and
shift resources where they are needed most, {\em importance sampling}
is used.

\subsection{Importance sampling}

When the variance of the function to be integrated is large, the error
[directly dependent on the variance, see Eq.~(\ref{eq:variance})] is
also large. A cure to the problem is provided by generating random
numbers that more efficiently sample the area, i.e., distributed
according to a function $p(x)$ which, if possible, has to fulfill the
following criteria: First, $p(x)$ should be as close as possible to
$f(x)$ and second, generating $p$-distributed random numbers should
be easily accomplished. The integral of $f(x)$ can be expressed in the
following way [using the notation introduced in Eq.~(\ref{eq:moments})]
\begin{equation}
\langle f \rangle 
= \langle f/p \rangle_p
=\int_0^1\frac{f(x)}{p(x)} p(x) dx
\approx\frac{1}{M} \sum_i^M \frac{f(y_i)}{p(y_i)}
\;.
\label{eq:imp}
\end{equation}
In Eq.~(\ref{eq:imp}) $\langle \cdots \rangle_p$ corresponds to a
sampling with respect to $p$-distributed random numbers; $y_i$ are
also $p$-distributed. The advantage of this approach is that the error
is now given in terms of the variance ${\rm Var} (f/p)$ and, if both
$f(x)$ and $p(x)$ are close, the variance of $f/p$ is considerably 
smaller than the variance of $f$.

For the case of $f(x) = x^n$ we could, for example, select random
numbers distributed according to $p(x) \sim x^\ell$ with $\ell \ge n$ (when
$n > -1$). This means that in Fig.~\ref{fig:reject} the area around $x
\lesssim 1$ is sampled with a higher probability than the area around
$x \sim 0$. Power-law distributed random numbers $y$ can be readily
produced from uniform random numbers $x$ by inverting the cumulative
distribution of $p(x)$, i.e., 
\begin{equation}
y(x) = x^{1/(\ell+1)}\;,
\;\;\;\;\;\;\;\;
\ell > -1\;.
\end{equation}

In the next sections the elaborated concepts are applied to problems in
(statistical) physics. First, some toy models and physical approaches
to study the critical behavior of statistical models using finite-size
simulations are introduced.

\section{Interlude: Statistical mechanics}

In this section the core concepts of statistical mechanics are
presented as well as a simple model to study phase transitions.
Because discussing these topics at length is beyond the scope
of these lecture notes, the reader is referred to the vast literature
in statistical physics, in particular Refs.~\cite{yeomans:92,huang:87,
goldenfeld:92,hartmann:01,cardy:96,stanley:71,reichl:98}.

\subsection{Simple toy model: The Ising model}

\begin{SCfigure}[][!htb]
  \includegraphics[scale=1.5]{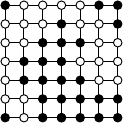}\hspace{2pc}
  \caption
   {
    Illustration of the two-dim\-ensional Ising model with nearest-neighbor
    interactions. Filled [open] circles represent $S_i = +1$ [$S_i = -1$].
    The spins only interact with their nearest neighbors (lines connecting
    the dots).
   }
  \label{fig:ising}
\end{SCfigure}

Developed in 1925 \cite{ising:25} by Ernst Ising and Wilhelm Lenz,
the Ising model has become over the decades the drosophila of
statistical mechanics.  The simplicity yet rich behavior of the
model makes it the perfect platform to study many magnetic systems
as well as for testing of algorithms. For simplicity, it is assumed
that the magnetic moments are highly anisotropic, i.e., they can
only point in one space direction. The classical spins $S_i = \pm 1$
are placed on a hypercubic lattice with nearest-neighbor interactions.
Therefore, the Hamiltonian is given by
\begin{equation}
{\mathcal H} = \sum_{\langle i,j\rangle}J_{ij}S_i S_j - H\sum_i S_i\; .
\label{eq:ising}
\end{equation}
The first term in Eq.~(\ref{eq:ising}) is responsible for the pairwise
interaction between two neighboring spins $S_i$ and $S_j$. When
$J_{ij} = -J <0$, the energy is minimized by aligning all spins, i.e.,
ferromagnetic order, whereas when $J_{ij} = J > 0$ the energy is
minimized by ensuring that the product over all neighboring spins
is negative. In this case, staggered antiferromagnetic order is
obtained for $T \to 0$. The ``$\langle i,j\rangle$'' represents a
sum over nearest-neighbor pairs of spins on the lattice
(see Fig.~\ref{fig:ising}).
The second term in Eq.~(\ref{eq:ising}) represents a coupling to
an external field of strength $H$. Amazingly, this simple model
captures all interesting phenomena found in the physics of statistical
mechanics and phase transitions. It is exactly solvable in one space
dimension, and in two dimensions for $H=0$, and thus an excellent test bed
for algorithms.
Furthermore, in space dimensions larger than one it undergoes a
finite-temperature transition into an ordered state. 

A natural way to quantify the temperature-dependent transition
in the ferromagnetic case is to measure the magnetization 
\begin{equation}
m = \frac{1}{N}\sum_i S_i
\end{equation}
of the system. When all spins are aligned, i.e.,
at low temperatures (below the transition), the magnetization is
close to unity. For temperatures much larger than the transition
temperature $T_c$, spins fluctuate wildly and so, on average, the
magnetization is zero.  Therefore, the magnetization plays the role
of an {\em order parameter} that is large in the ordered phase and
zero otherwise.  Before the model is described further, some basic
concepts from statistical physics are introduced.

\subsection{Statistical physics in a nutshell}

It would be beyond the scope of this lecture to discuss in detail
statistical mechanics of magnetic systems. The reader is referred
to the vast literature on the topic \cite{yeomans:92,huang:87,
goldenfeld:92,hartmann:01,cardy:96,stanley:71,reichl:98}. In
this context only the relevant aspects of statistical physics are
discussed.

\paragraph{Observables} In statistical physics, expectation values
of quantities such as the energy, magnetization, specific heat,
etc.---generally called {\em observables}---are computed by performing
a trace over the partition function ${\mathcal Z}$. Within the {\em
canonical ensemble} \cite{huang:87} where the temperature $T$ is
fixed, the {\em expectation value} or thermal average of an observable
${\mathcal O}$ is given by
\begin{equation}
\langle {\mathcal O}\rangle = \frac{1}{\mathcal Z}\sum_{s}{\mathcal
O}(s)e^{-{\mathcal H}(s)/kT}\;.
\label{eq:obs}
\end{equation}
The sum is over all states $s$ in the system, and $k$ represents
the Boltzmann constant. ${\mathcal Z} = \sum_{s}\exp[{-{\mathcal
H}(s)/kT}]$ is the partition function which normalizes the equilibrium
Boltzmann distribution
\begin{equation}
{\mathcal P}_{\rm eq}(s) = \frac{1}{\mathcal Z} e^{-{\mathcal H}(s)/kT}\;.
\label{eq:boltz}
\end{equation}
The $\langle \cdots \rangle$ in Eq.~(\ref{eq:obs}) represent a thermal
average. One can show that the internal energy of the system is given
by
\begin{equation}
E = \langle {\mathcal H}(s)\rangle\;,
\label{eq:eint}
\end{equation}
whereas the free energy ${\mathcal F}$ is given by
\begin{equation}
{\mathcal F} = -kT \ln {\mathcal Z}\;.
\label{eq:efree}
\end{equation}
Note that {\em all} thermodynamic quantities can be computed directly
from the partition function and expressed as derivatives of the
free energy (see Ref.~\cite{yeomans:92} for details). Because the
partition function is closely related to the Boltzmann distribution,
it follows that if we can {\em sample} observables (e.g., measure the
magnetization) with states generated according to the corresponding
Boltzmann distribution, a simple Markov-chain ``integration'' scheme
can be used to produce an estimate.

\paragraph{Phase transitions} Continuous phase transitions
\cite{huang:87} have no latent heat at the transition and are thus
easier to describe. At a continuous phase transition the free energy
has a singularity that usually manifests itself via a power-law behavior of
the derived observables at criticality. The correlation length $\xi$
\cite{huang:87}---which gives us a measure of correlations and order
in a system---diverges at the transition 
\begin{equation}
\xi \sim |T - T_c|^{-\nu}\;,
\end{equation}
with $\nu$ a critical exponent quantifying this divergence and $T_c$
the transition temperature. Close enough to the transition (i.e.,
$|T - T_c|/T_c \ll 1$) the behavior of observables can be well
described by power laws. For example, the specific heat $c_V$ has a
singularity at $T_c$ with $c_V \sim |T - T_c|^{-\alpha}$, although the
exponent $\alpha$ (unlike $\nu$) can be both negative and positive. The
magnetization does not diverge, but has a singular kink at $T_c$, i.e.,
$m \sim |T - T_c|^\beta$ with $\beta > 0$. 

Using arguments from the renormalization group \cite{goldenfeld:92}
it can be shown that the critical exponents are related via {\em
scaling} relations. Often (as in the Ising case), only two exponents
are independent and
{\em fully} characterize the critical behavior of the model.
It can be further shown that models in statistical physics generally obey
{\em universal behavior} (there are some exceptions\ldots), i.e.,
if the lattice geometry is kept the same, the critical exponents
{\em only} depend on the order parameter symmetry.  Therefore, when
simulating a statistical model, it is enough to determine the location
of the transition temperature $T_c$, as well as {\em two independent}
critical exponents to fully characterize the {\em universality class}
of the system.

\paragraph{Finite-size scaling and the Binder ratio (or ``Binder cumulant'')} How can we
determine the {\em bulk} critical exponents of a system by simulating
finite lattices? When the systems are not infinitely large, the
critical behavior is smeared out.  Again, using arguments from the
renormalization group, one can show that the nonanalytic part of
a given observable can be described by a {\em finite-size scaling}
form \cite{privman:90}. For example, the finite-size magnetization
from a simulation of an Ising system with $L^d$ spins is asymptotically
(close to the transition, and for large $L$) given by
\begin{equation}
\langle m_L \rangle \sim L^{\beta/\nu}\tilde{M}[L^{1/\nu}(T - T_c)]\;,
\label{eq:fssm}
\end{equation}
and for the magnetic susceptibility by
\begin{equation}
\chi_L \sim L^{\gamma/\nu}\tilde{C}[L^{1/\nu}(T - T_c)]\;,
\label{eq:fsss}
\end{equation}
where close to the transition $\chi \sim |T - T_c|^{-\gamma}$ 
(for the infinite system, $L\to\infty$) and
\begin{equation}
\chi = \frac{L^d}{kT} \left(\langle m^2 \rangle - \langle m\rangle^2\right)\;.
\end{equation}
Both $\tilde{M}$ and $\tilde{C}$ are unknown {\em scaling functions}.
Equations (\ref{eq:fssm}) and (\ref{eq:fsss}) show that when $T =
T_c$, data for $\langle m_L \rangle / L^{\beta/\nu}$ and $\chi_L /
L^{\gamma/\nu}$ simulated for different system sizes $L$ should cross
in the large-$L$ limit at one point, namely $T = T_c$, provided we use
the right expressions for $\beta/\nu$ and $\gamma/\nu$, respectively.
In reality, there are nonanalytic corrections to scaling and so
the crossing points between two {\em successive} system size pairs
(e.g., $L$ and $2L$) converge to a common crossing point for $L \to
\infty$ that agrees with the bulk transition temperature $T_c$.
Performing the finite-size scaling analysis with the magnetization or
the susceptibility is not very practical, because neither $\beta$ nor
$\gamma$ are known a priori. There are other approaches to determine
these, but a far simpler method is to determine {\em combined}
quantities that are dimensionless. One such quantity is known as the
{\em Binder ratio} (or ``Binder cumulant'') \cite{binder:81} given by
\begin{equation}
g = \frac{1}{2}\left[
3 - \frac{\langle m^4\rangle}{\langle m^2\rangle^2}
\right] \sim
\tilde{G}[L^{1/\nu}(T - T_c)]\;.
\label{eq:binder}
\end{equation}
The different factors ensure that $g \to 1$ for $T \to 0$ and $g
\to 0$ for $T \to \infty$.
The asymptotic (for large $L$) scaling behavior of the Binder ratio
follows directly from the fact that the pre-factors
of the moments of the magnetization ($m^k \sim
L^{k\beta/\nu}$) cancel out in Eq.~(\ref{eq:binder}).  

\begin{figure}[!htb]
\centering
\includegraphics[scale=0.30]{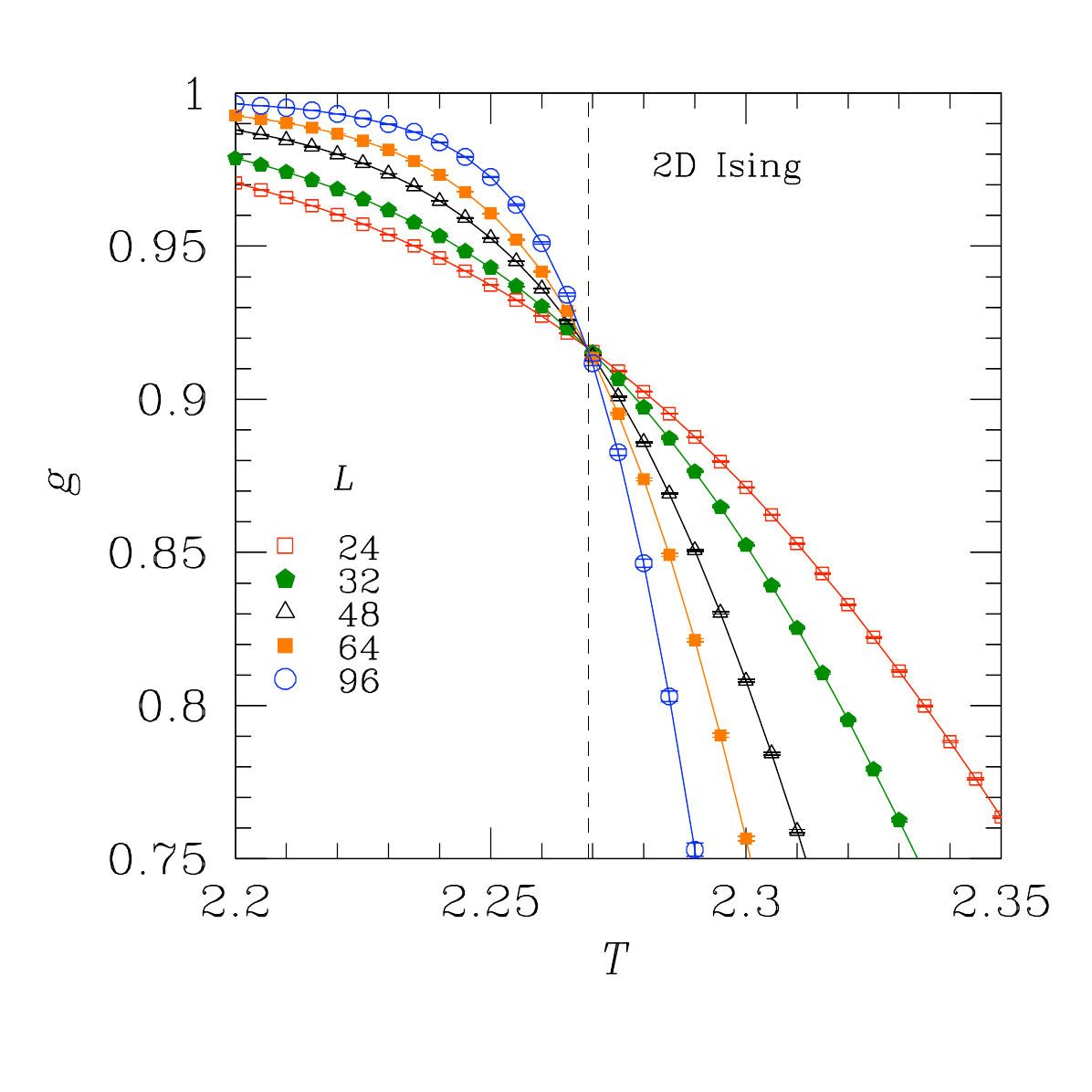}
\includegraphics[scale=0.30]{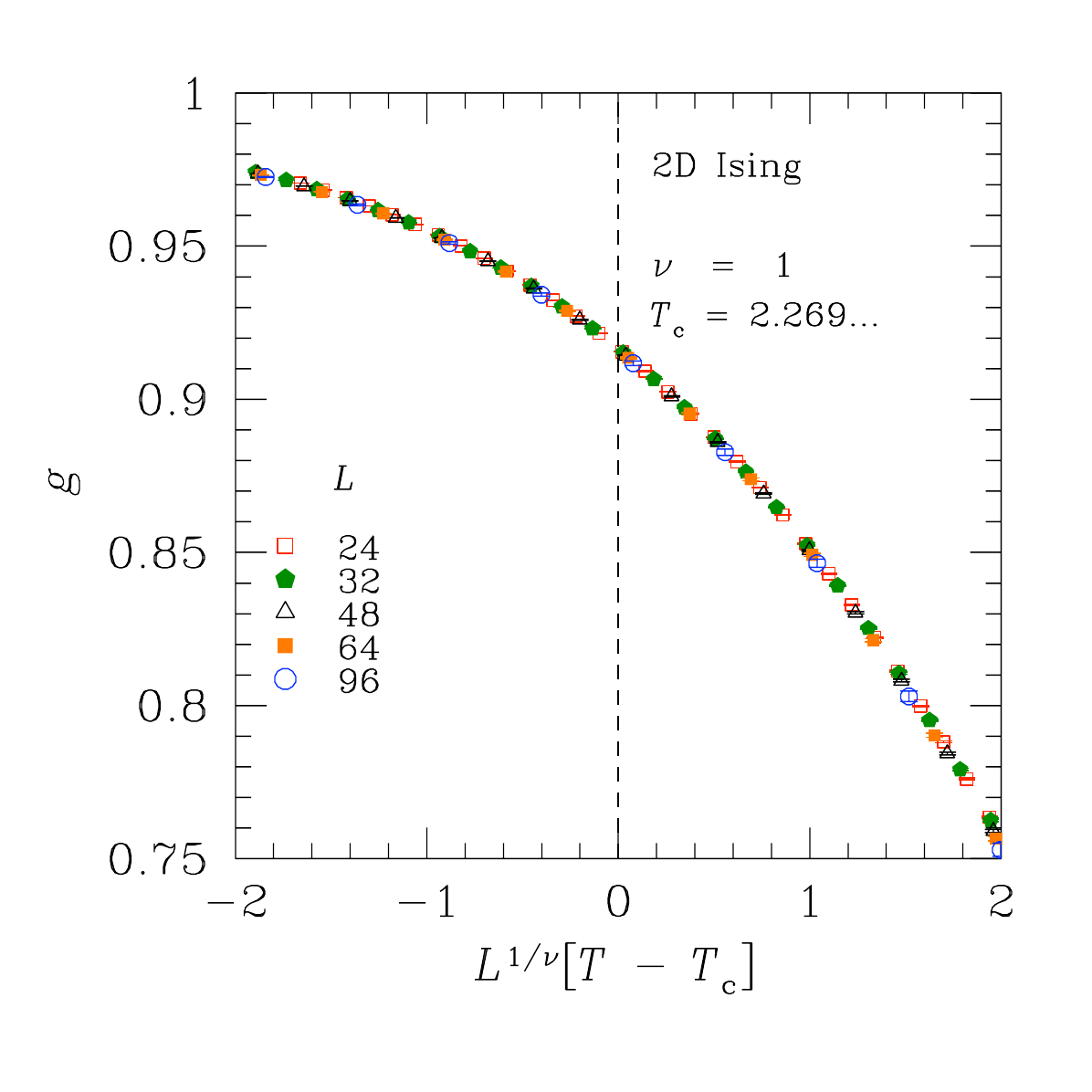}\vspace{-2pc}
\caption
{
Left panel: Binder ratio as a function of temperature for the
two-dimensional Ising model with nearest-neighbor interactions. The data
approximately cross at one point (the dashed line corresponds
to the exactly-known $T_c$
for the two-dimensional Ising model) signaling a transition. Right
panel: Finite-size scaling of the data in the left panel using the
known $T_c = 2.269\ldots$ and $\nu = 1$.  Plotted are data for the
Binder ratio as a function of the scaling variable $L^{1/\nu}[T -
T_c]$.  Data for different system sizes fall onto a universal curve
suggesting that the parameters used are the correct ones.
}
\label{fig:isingmc}
\end{figure}

The Binder ratio is a {\em dimensionless} quantity and so data for
different system sizes $L$ approximately cross at a putative
transition---provided corrections to scaling are small.
Furthermore,
by carefully selecting the correct value of the critical exponent
$\nu$, the data fall onto a universal curve. Therefore, the method
allows for an estimation of $T_c$, as well as the critical
exponent $\nu$.  This is illustrated in Fig.~\ref{fig:isingmc} for the
two-dimensional Ising model. The left panel shows the Binder ratio
as a function of temperature for several small system sizes. The
vertical dashed line marks the exactly-known value of the critical
temperature, namely $T_c = 2/\ln(1 + \sqrt{2}) \approx 2.269\ldots$
\cite{yeomans:92}.  The right panel shows a finite-size scaling
analysis of the data for the exact value of the critical exponent
$\nu$. Close to the transition the data fall onto a universal
curve, showing that $\nu = 1$ is the correct value of the critical
exponent. The two-dimensional Ising universality class is fully
characterized with a second critical exponent, e.g., $\beta = 1/8$.

Note that other dimensionless quantities, such as the two-point
finite-size correlation length \cite{palassini:99b,ballesteros:00}
can also be used with similar results.

\section{Monte Carlo simulations in statistical physics}

In analogy to the importance-sampling Monte Carlo integration of
functions discussed in Sec.~\ref{sec:integration}, we can use the gained
insights to sample the average of an observable in statistical physics.
In general, as shown in Eq.~(\ref{eq:obs}),
\begin{equation}
\langle {\mathcal O} \rangle = 
\frac{\sum_{s}{\mathcal O}(s)e^{-{\mathcal H}(s)/kT}}{\sum_{s}{e^{-{\mathcal H}(s)/kT}}}\;.
\label{eq:obs2}
\end{equation}
Equation (\ref{eq:obs2}) can be trivially extended with a distribution for 
the states, i.e., 
\begin{equation}
\langle {\mathcal O} \rangle = 
\frac{\sum_{s}[{\mathcal O}(s)e^{-{\mathcal H}(s)/kT}/{\mathcal P}(s)]
\,{\mathcal P}(s)}{\sum_{s}[e^{-{\mathcal H}(s)/kT}/{\mathcal P}(s)]
\,{\mathcal P}(s)}\;.
\label{eq:obs3}
\end{equation}
The approach is completely analogous to the importance sampling Monte
Carlo integration.  If ${\mathcal P}(s)$ is the Boltzmann distribution
[Eq.~(\ref{eq:boltz})] then the factors cancel out and we obtain
\begin{equation}
\langle {\mathcal O} \rangle = \frac{1}{M}\sum_i {\mathcal O}(s_i)\;, 
\end{equation}
where the states $s_i$ are now selected according to the Boltzmann
distribution. The problem now is to find an algorithm that
allows for a sampling of the Boltzmann distribution. The method is
known as the Metropolis algorithm.

\subsection{Metropolis algorithm}

The Metropolis algorithm \cite{metropolis:53} was developed in 1953
at Los Alamos National Lab within the nuclear weapons program mainly
by the Rosenbluth and Teller families \cite{comment:mc}. The article 
in the Journal of Chemical Physics starts in the following way:

\begin{center}
\begin{minipage}[b]{28pc} 
``{\em The purpose of this paper is to describe a general method, suitable
for fast electronic computing machines, of calculating the properties
of any substance which may be considered as composed of interacting
individual molecules.}''
\end{minipage}
\end{center}

\noindent And they were right. The idea is the following: In order to
evaluate Eq.~(\ref{eq:obs2}) we generate a Markov chain of successive
states $s_1 \to s_2 \to \ldots$. The new state is generated from
the old state with a carefully-designed transition probability ${\mathcal P}(s
\to s')$ such that it occurs with a probability given by the
equilibrium Boltzmann distribution, i.e., ${\mathcal P}_{\rm eq}(s) =
Z^{-1}\exp[-{\mathcal H}(s)/kT]$. In the Markov process, the state $s$
occurs with probability ${\mathcal P}_k(s)$ at the $k$th time step,
described by the master equation
\begin{equation}
{\mathcal P}_{k+1}(s) = {\mathcal P}_k(s) + \sum_{s'} 
\left[
\mathcal{T}(s' \to s){\mathcal P}_k(s')
-
\mathcal{T}(s \to s'){\mathcal P}_k(s)
\right]\;.
\label{eq:master}
\end{equation}
The sum is over all states $s'$ and the first term in the sum describes
all processes {\em reaching} state $s$, while the second term describes
all processes {\em leaving} state $s$. The goal is that for $k \to
\infty$ the probabilities ${\mathcal P}_k(s)$ reach a stationary
distribution described by the Boltzmann distribution. The transition
probabilities $\mathcal{T}$ can be designed in such a way that for ${\mathcal
P}_k(s) = {\mathcal P}_{\rm eq}(s)$, all terms in the sum vanish,
i.e., for all $s$ and $s'$ the {\em detailed balance} condition
\label{page:detbal}
\begin{equation}
\mathcal{T}(s' \to s){\mathcal P}_{\rm eq}(s')
=
\mathcal{T}(s \to s'){\mathcal P}_{\rm eq}(s)
\label{eq:db}
\end{equation}
must hold. The condition in Eq.~(\ref{eq:db}) means that the
process has to be reversible. Furthermore, when the system has assumed
the equilibrium probabilities, the ratio of the transition probabilities
only depends on the change in energy $\Delta {\mathcal H}(s,s') = {\mathcal
H}(s') - {\mathcal H}(s)$, i.e., 
\begin{equation}
\frac{\mathcal{T}(s \to s')}{\mathcal{T}(s' \to s)} = \exp[-({\mathcal H}(s') - {\mathcal H}(s))/kT] = \exp[-\Delta {\mathcal H}(s,s')/kT]\;.
\label{eq:fraq}
\end{equation}
There are different choices for the transition probabilities $\mathcal{T}$
that satisfy Eq.~(\ref{eq:fraq}). One can show that $\mathcal{T}$ has to
satisfy the general equation $\mathcal{T}(x)/\mathcal{T}(1/x) = x$ $\forall x$ with
$x = \exp(-\Delta{\mathcal H}/kT)$. There are two convenient choices
for $\mathcal{T}$ that satisfy this condition:

\paragraph{1. Metropolis (also known as Metropolis-Hastings) algorithm} In this case $\mathcal{T}(x) = \min(1,x)$ and so
\begin{equation}
\mathcal{T}(s \to s') =  
\left\{ 
\begin{array}{ll}
         \Gamma, & \mbox{if $\Delta{\mathcal H} \le 0$};\\
         \Gamma e^{-\Delta{\mathcal H}(s,s')/kT}, & \mbox{if $\Delta{\mathcal H} \ge 0$}\;.
\end{array} 
\right.
\label{eq:metro}
\end{equation}
In Eq.~(\ref{eq:metro}), $\Gamma^{-1}$ represents a Monte Carlo time.

\paragraph{2. Heat-bath algorithm} In this case $\mathcal{T}(x) =
x/(1+x)$ corresponding to an acceptance probability $\sim [1+
\exp(\Delta{\mathcal H}(s,s')/kT)]^{-1}$.  For the rest of this lecture,
we focus on the Metropolis algorithm.  The heat bath algorithm is
more efficient when temperatures far below the transition temperature
are sampled.

The move between states $s$ and $s'$ can, in principle, be
arbitrary. If, however, the energies of states $s$ and $s'$ are too
far apart, the move will likely not be accepted. For the case of the
Ising model, in general, a single spin $S_i$ is selected and flipped
with the following probability:
\begin{equation}
\mathcal{T}(S_i \to -S_i) =  
\left\{ 
\begin{array}{ll}
         \Gamma, & \mbox{for $S_i = -{\rm sign}(h_i)$};\\
         \Gamma e^{-2S_ih_i/kT}, & \mbox{for $S_i = {\rm sign}(h_i)$}\;.
\end{array} 
\right.
\end{equation}
where $h_i = -\sum_{j\neq i}J_{ij}S_j + H$ is the effective local field
felt by the spin $S_i$.

\newpage

\paragraph{Practical implementation of the Metropolis algorithm}
A simple pseudo-code Monte Carlo program to compute an observable
\texttt{O} for the Ising model is the following:

\SourceCodeLines{99}
\begin{Verbatim}[fontsize=\small]
 algorithm ising_metropolis(T,steps)
     initialize starting configuration S
     initialize O = 0

     for(counter = 1 ... steps) do
         generate trial state S'
         compute p(S -> S',T)
         x = rand(0,1)
         if(p > x) then
             accept S' 
         fi

         O += O(S')
     done

 return  O/steps
\end{Verbatim}

\noindent After initialization, in line 6 a proposed state is generated
by, e.g., flipping a spin. The energy of the new state is computed
and henceforth the transition probability between states \texttt{p}
$= \mathcal{T}(S\to S')$. A uniform random number \texttt{x} $\in [0,1]$ is
generated. If the probability is larger than the random number, the
move is accepted. If the energy is lowered, i.e., $\Delta{\mathcal H}
> 0$, the spin is always flipped. Otherwise the spin is flipped with a
probability \texttt{p}. Once the new state is accepted, we measure a
given observable and record its value to perform the thermal average
at a given temperature. For \texttt{steps} $\to \infty$ the average
of the observable converges to the exact value, again with an error
inversely proportional to the square root of the number of steps.
This is the core bare-bones routine for the Metropolis algorithm. In
practice, several aspects have to be considered to ensure that the
data produced are correct. The most important, autocorrelation and
equilibration times, are described below.

\subsection{Equilibration} In order to obtain a correct estimate
of an observable ${\mathcal O}$, it is imperative to ensure that
one is actually sampling an {\em equilibrium} state. Because, in
general, the initial configuration of the simulation can be chosen at
random---popular choices being random or polarized configuration---the
system will have to evolve for several Monte Carlo steps before
an equilibrium state at a given temperature is obtained. The time
$\tau_{\rm eq}$ until the system is in thermal equilibrium is called
{\em equilibration time} and depends directly on the system size
(e.g., the number of spins $N = L^d$) and increases with decreasing
temperature. In general, it is measured in units of {\em Monte Carlo
sweeps} (MCS), i.e., $1$ MCS $= N$ spin updates.

\begin{figure}[!htb]
\centering
\includegraphics[scale=0.50]{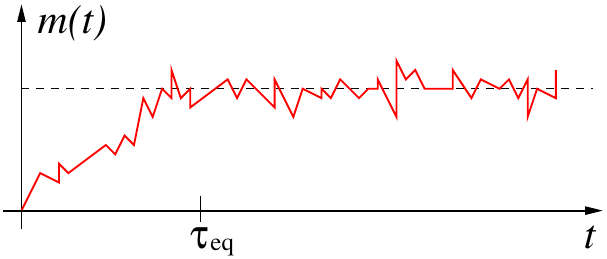}
\caption
{
Sketch of the equilibration behavior of the magnetization $m$ as a
function of Monte Carlo time. After a certain time $\tau_{\rm eq}$
the data become approximately flat and fluctuate around a mean value.
Once $\tau_{\rm eq}$ has been reached, the system is in thermal
equilibrium and observables can be measured.
}
\label{fig:equil}
\end{figure}

\noindent In practice, all measured observables should be monitored
as a function of MCS to ensure that the system is in thermal
equilibrium. Some observables, such as the energy, equilibrate faster
than others (e.g., magnetization) and thus the equilibration times
of {\em all} observables measured need to be considered.

\subsection{Autocorrelation times and error analysis}

Because in a Markov chain the new states are generated by modifying
the previous ones, subsequent states can be highly correlated. To
ensure that the measurement of an observable ${\mathcal O}$ is not
biased by correlated configurations, it is important to measure the
autocorrelation time $\tau_{\rm auto}$ that describes the time it takes
for two measurements to be decorrelated. This means that in a Monte
Carlo simulation, after the system has been thermally equilibrated,
measurements can only be taken every $\tau_{\rm auto}$ MCS. To compute
the autocorrelation time for a given observable ${\mathcal O}$,
the time-dependent {\em autocorrelation} function needs to be measured:
\begin{equation}
C_{\mathcal O}(t) = \frac{\langle {\mathcal O}(t_0){\mathcal O}(t_0+ t)\rangle
- \langle {\mathcal O}(t_0)\rangle\langle {\mathcal O}(t_0 + t)\rangle}{
\langle {\mathcal O}^2(t_0)\rangle - \langle {\mathcal O}(t_0)\rangle^2}\;.
\label{eq:cauto}
\end{equation}
In general, $C_{\mathcal O}(t) \sim \exp(-t/\tau_{\rm auto})$ and
so $\tau_{\rm auto}$ is given by the value where $C_{\mathcal O}$
drops to $1/e$. An alternative is the integrated autocorrelation
time $\tau_{\rm auto}^{\rm int}$. It is basically the same as the
standard autocorrelation time for any practical purpose. However,
it is easier to compute:
\begin{equation}
\tau_{\rm auto}^{\rm int} = \frac{\sum_{t = 1}^\infty \left(\langle {\mathcal O}(t_0){\mathcal O}(t_0+ t)\rangle -  \langle {\mathcal O}\rangle^2\right)}{\langle {\mathcal O}^2\rangle - \langle {\mathcal O}\rangle^2}
\end{equation}
Autocorrelation effects influence the determination of the error of
statistical estimates. It can be shown \cite{hartmann:01} that the
error $\Delta {\mathcal O}$ is given by
\begin{equation}
\Delta {\mathcal O} = 
\sqrt{
\frac{\langle {\mathcal O}^2\rangle - \langle {\mathcal O}\rangle^2}{(M - 1)}(1 + 2\tau_{\rm auto})\;.
}
\end{equation}
Here $M$ is the number of measurements.  The autocorrelation
time directly influences the calculation of the error bars and must
be computed and included in all calculations. So far, we have not
discussed how the autocorrelation times depend on the system size and
the temperature. Like the equilibration times, the autocorrelation
times increase with increasing system size.

\subsection{Critical slowing down and the Wolff cluster algorithm} 
\label{subsec:wolff}

Close to a phase transition, the autocorrelation time is given by
\begin{equation}
\tau_{\rm auto} \sim \xi^z
\end{equation}
with $z > 1$ and typically around $2$. Because the correlation
length $\xi$ diverges at a continuous phase transition, so does the
autocorrelation time. This effect, known as {\em critical slowing
down}, slows simulations to intractable times close to continuous
phase transitions when the {\em dynamical critical exponent} $z$
is large.  

The problem can be alleviated by using Monte Carlo methods which,
while only performing small changes to the energy of the system
(to ensure that moves are accepted frequently), heavily randomize
the spin configurations and not only change the value of one spin.
This ensures that phase space is sampled evenly.  Typical examples
are {\em cluster algorithms} \cite{swendsen:87,wolff:89} where a
carefully-built cluster of spins is flipped at each step of the
simulation \cite{landau:97,landau:00,newman:99,hartmann:01}.

\paragraph{Wolff cluster algorithm (Ising spins)} In the Wolff cluster
algorithm \cite{wolff:89} we choose a random spin and build a cluster
around it (the algorithm is constructed in such a way that larger
clusters are preferred). Once the cluster is constructed, it is flipped
in a rejection-free move. This ``randomizes'' the system efficiently,
thus overcoming critical slowing down.  Outline of the algorithm:

\begin{itemize}

\item[$\Box$]{Select a random spin.}

\item[$\Box$]{If a neighboring spin is parallel to the initial spin,
add it to the cluster with a probability $1 - \exp(-2J/kT)$.}

\item[$\Box$]{Repeat the previous step for all neighbors of the
newly-added spins and iterate until no new spins can be added.}

\item[$\Box$]{Flip all spins in the cluster.}

\end{itemize}

\noindent The algorithm obeys detailed balance. Furthermore, one
can show that the linear size of the cluster is proportional to the
correlation length. Therefore the algorithm adapts to the behavior
of the system at criticality resulting in $z \approx 0$, i.e., the
critical slowing down encountered around the transition is removed
and the algorithm performs orders of magnitude faster than simple
Monte Carlo. For low temperatures, the cluster algorithm merely
``flip-flops'' almost all spins of the system and provides not
much improvement, unless a domain wall is stuck in the system. For
temperatures much higher than the critical temperature the size of
the clusters is of the order of one spin and there the Metropolis
algorithm outperforms the cluster algorithm (keep in mind that
building the cluster takes many operations). Thus the method works
best at criticality.

In general, to be able to cover a temperature range that extends
beyond the critical region, combinations of cluster updates and
local updated (standard Monte Carlo) are recommended.  One can
also define {\em improved estimators} to measure observables with
a reduced statistical error. Finally, the Wolff cluster algorithm
can also be generalized to Potts spins, XY and Heisenberg spins,
as well as hard disks. The reader is referred to the literature
\cite{landau:97,landau:00,newman:99,hartmann:01,krauth:06} for details.

Note that the Swendsen-Wang cluster algorithm \cite{swendsen:87} is
similar to the Wolff cluster algorithm. However, instead of building
one cluster, multiple clusters are built. This is less efficient when
the space dimension is larger than two because in that case only few
large clusters will exist.

\subsection{When does simple Monte Carlo fail?}

\begin{SCfigure}[1.1][!htb]
  \includegraphics[scale=0.40]{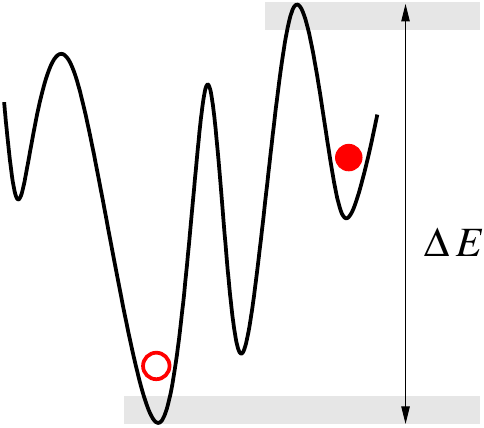}\vspace{1.5pc}\hspace{2pc}
  \caption
   {
    Sketch of a rough energy landscape. A Monte Carlo move from the 
    initial (solid circle) to the final state (open circle) is unlikely
    if the size of the barrier $\Delta E$ is large, especially at low
    temperatures. A simple Monte Carlo simulation will stall and the
    system will be stuck in the metastable state.
   }
  \label{fig:land}
\end{SCfigure}

Metropolis {\em et al.}~did not bear in mind that there are systems
where even a simple spin flip can produce a huge change in the energy
$\Delta {\mathcal H}$ of the system. This has the effect that the
probability for new configurations to be accepted is very small and
the simulation stalls, in particular when the studied system has a
rough energy landscape, i.e., different states in phase space are
separated by large energy ``mountains'' and deep energy ``valleys,''
as depicted in Fig.~\ref{fig:land}.  Examples of such complex systems
are spin glasses, proteins and neural networks.

\noindent These systems are characterized by a complex energy
landscape with deep valleys and mountains that  grow exponentially
with the system size.  Therefore, for low temperatures, equilibration
times of simple Monte Carlo methods diverge.  Although the method
technically still works, the time it takes to equilibrate even the
smallest systems becomes impractical. Improved sampling techniques
for rough energy landscapes need to be implemented.

\section{Complex toy model: The Ising spin glass}
\label{sec:sg}

\begin{figure}[!htb]
\centering
\includegraphics[scale=1.50]{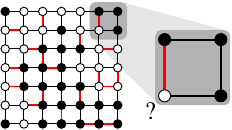}
\caption
{
Two-dimensional Ising spin-glass. The circles represent Ising spins.
A thin line between two spins $i$ and $j$ corresponds to $J_{ij} < 0$,
whereas a thick line corresponds to $J_{ij} > 0$.  In comparison to
a ferromagnet, the behavior of the model system changes drastically,
as illustrated in the highlighted plaquette. For $T \to 0$, the spin
in the lower left corner is unable to fulfill the interactions with
the neighbors and is {\em frustrated} (see text).
}
\label{fig:sg}
\end{figure}

What happens if we take the ferromagnetic Ising model and flip the
sign of randomly-selected interactions $J_{ij}$ between two spins?
The resulting behavior is illustrated in Fig.~\ref{fig:sg}. For
low temperatures, if the product of the interactions $J_{ij}$
around any plaquette is negative, {\em frustration} effects
emerge. The spin in the lower left corner of the highlighted
plaquette tries to minimize the energy by either aligning with the
right neighbor, or being antiparallel with the top neighbor. Both
conditions are mutually exclusive and so the energy cannot be
uniquely minimized.  This behavior is a hallmark of spin glasses
\cite{binder:86,mezard:87,young:98,fisher:91c,
talagrand:03,diep:05,dedominicis:06}. Note that, in general, the
bonds are either chosen from a bimodal (${\mathcal P}_b$) or Gaussian
(${\mathcal P}_g$) disorder distribution:
\begin{equation}
{\mathcal P}_b(J_{ij}) = p\delta(J_{ij} - 1) + (1-p)\delta(J_{ij} + 1)\;,
\;\;\;\;\;\;
{\mathcal P}_g(J_{ij}) = \frac{1}{\sqrt{2\pi}}\exp[-J_{ij}^2/2]\;,
\end{equation}
where, in general $p = 1/2$. The Hamiltonian in Eq.~(\ref{eq:ising})
with disorder in the bonds is known as the Edwards-Anderson Ising spin
glass. There is a finite-temperature transition for space dimensions
$d \ge 3$ between a spin-glass and the (thermally) disordered state,
cf.~Sec.~\ref{subsec:spinglasstheory}.
For example, for Gaussian disorder in three space dimensions
$T_c \approx 0.95$ \cite{katzgraber:06}.

The result of competing interactions is a complex energy landscape. The
complexity of the model increases considerably. For example, finding
the ground state energy of a spin glass is generally an NP-hard
problem.  Equilibration times in finite-temperature Monte Carlo
simulations grow exponentially and thus the study of system sizes
beyond a few spins becomes intractable. So \ldots why study these
systems? Not only are there many materials \cite{binder:86} that can
be described well with spin-glass Hamiltonians, many other problems
spanning several fields of science can be either described directly by
spin-glass Hamiltonians or mapped onto these. Therefore these models
are of general interest to a broad spectrum of disciplines.

Note that, because in general only finite system sizes can be
simulated, an average over different realizations of the disorder
needs to be performed in addition to the thermal averages. This
means that after a Monte Carlo simulation has been completed for
a given distribution of the disorder, it must be repeated at least
$1000$ times for the results to be representative.  Although this
extra effort further complicates simulations of spin-glass systems,
it makes them {\em embarrassingly} parallel, i.e., simulations can
easily be distributed over many workstations.

\subsection{Selected hallmark properties of spin glasses} 

Because of the complex energy landscape, spin glasses show dynamical
properties not seen in any other materials/systems. First, spin-glass
observables such as susceptibilities and magnetizations {\em age}
with time. Due to the complex energy landscape, there are rearrangements
of the spins in macroscopic time scales. Therefore, when preparing a
spin-glass system at a given temperature, a slow decay of observables
can be observed because the system, at least experimentally,
is never in thermal equilibrium. Furthermore, when performing an
aging experiment on a spin glass, changing the temperature from
$T_1 < T_c$ to $T_2 < T_1$ at time $t_1$ for a finite period of time
$t_2$ and then back to $T_1$ shows interesting {\em memory} effects
\cite{jonason:98}. After the time $t_1 + t_2$, the system remembers the
state it had at time $t_1$ and follows the previous aging path. This
memory and rejuvenation effect is unique to spin glasses.

While the susceptibility shows a cusp at the transition
\cite{cannella:72}, the specific heat has a smooth behavior around
the transition temperature \cite{mezard:87}.  Furthermore, no
signs of spatial ordering can be found when performing a neutron
scattering experiment probing below the transition temperature.
However, M\"ossbauer spectroscopy shows that the magnetic moments are
frozen in space, thus indicating that the system is in a glassy and
not liquid phase. Therefore, in its simplest interpretation, a spin
glass is a model for a highly-disordered magnet.

\subsection{Theoretical description}
\label{subsec:spinglasstheory}

In 1975, Edwards and Anderson suggested a phenomenological model in
order to describe these fascinating materials: the Edwards-Anderson
(EA) spin-glass Hamiltonian \cite{edwards:75} discussed above.
In 1979, Parisi postulated a solution (only recently proven to be
correct \cite{talagrand:06}) to the mean-field Sherrington-Kirkpatrick
(SK) model \cite{sherrington:75}, a variation of the Edwards-Anderson
model with {\em infinite-range interactions} (all spins interact
with each other). The replica symmetry breaking picture (RSB) of
Parisi for the mean-field model spawned an increased interest in the
field and has been applied to a variety of problems. In addition, in
1986 a phenomenological description, called the ``Droplet Picture''
(DP) was introduced simultaneously by Fisher \& Huse and Bray \&
Moore \cite{fisher:86,fisher:87} in order to describe short-range
spin glasses, as well as the chaotic pairs picture by Newman \&
Stein \cite{newman:96,newman:07}  However, rigorous analytical
results are difficult to obtain for realistic short-range spin-glass
models. Because of this, research has shifted to intense {\em
numerical studies}.

Nevertheless, spin glasses are far from being understood. The
memory effect in spin glasses \cite{jonason:98} has yet to
be understood theoretically, and only recently was it observed
numerically \cite{jimenez:05}. The existence of a spin-glass phase
in a finite magnetic field \cite{dealmeida:78} has been a source
of debate \cite{young:04}, as well as the ultrametric structure of
the phase space (hierarchical structure of states) which remains
to be understood for realistic models \cite{hed:03,katzgraber:09}.
Finally, there have been several numerical attempts at finite
\cite{krzakala:00,palassini:00,marinari:00,katzgraber:01,houdayer:00}
and zero temperature \cite{hartmann:99,hartmann:01a} to
better understand the nature of the spin-glass state for short-range
spin glasses. To date, the data for Ising spins are consistent with
an intermediate picture \cite{krzakala:00,palassini:00} that combines
elements from the standard theoretical predictions of replica symmetry
breaking and the droplet theory.

How can order be quantified in a system that intrinsically does
not have visible spatial order? For this we need to first determine
what differentiates a spin glass at temperatures above the critical
point $T_c$ and below. Above the transition, like for the regular
Ising model, spins fluctuate and any given snapshot yields a random
configuration.  Therefore, comparing a snapshot at time $t$ and time
$t + \delta t$ yields completely different results. Below the
transition, (replica) symmetry is broken and configurations freeze
into place. Therefore, comparing a snapshot of the system at time
$t$ and time $t + \delta t$ shows significant similarities.
A natural choice thus is to define an {\em overlap} function $q$
which compares two copies of the system with the {\em same} disorder.

\begin{figure}[!htb]
\centering
\includegraphics[scale=1.00]{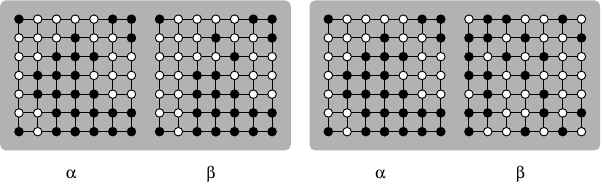}
\caption
{
Graphical representation of the order parameter function $q$. Two
replicas of the system $\alpha$ and $\beta$ with the same disorder
are compared spin-by-spin. The left set corresponds to a temperature
$T \ll T_c$ where many spins agree and so $q \to 1$ (in the depicted
example $q = 0.918$). The right set corresponds to $T > T_c$; the
spins fluctuate due to thermal fluctuations and so $q < 1$ (here $q
= 0.408$).
}
\label{fig:q}
\end{figure}

In simulations, it is less practical to compare two snapshots of the
system at different times. Therefore, for practical reasons two copies
(called ``replicas'') $\alpha$ and $\beta$ with the same disorder but
{\em different} initial conditions and Markov chains are simulated
in parallel. The order parameter is then given by
\begin{equation}
q = \frac{1}{N}\sum_i S_i^\alpha S_i^\beta\;,
\end{equation}
and is illustrated graphically in Fig.~\ref{fig:q}. For temperatures
below $T_c$, $q$ tends to unity whereas for $T > T_c$ on average
$q \to 0$, similar to the magnetization for the Ising ferromagnet.
Analogous to the ferromagnetic case, we can define a Binder ratio $g$
by replacing the magnetization $m$ with the spin overlap $q$ to probe
for the existence of a spin-glass state.

\section{Parallel tempering Monte Carlo}

As illustrated with the case of spin glasses in Sec.~\ref{sec:sg},
the free energy landscape of many-body systems with competing phases
or interactions is generally characterized by many local minima that
are separated by free-energy barriers. The simulation of these systems
with standard Monte Carlo \cite{metropolis:49,landau:97,krauth:98} or
molecular dynamics \cite{frenkel:96} methods is slowed down by long
relaxation times due to the suppression of tunneling through these
barriers. Already simple chemical reactions with latent heat, i.e.,
first-order phase transitions, present huge numerical challenges
that are not present for systems which undergo second-order phase
transitions where improved updating techniques, such as cluster
algorithms \cite{swendsen:87,wolff:89}, can be used.  For complex
systems with competing interactions, one instead attempts to improve
the local updating technique by introducing artificial statistical
ensembles such that tunneling times through barriers are reduced and
autocorrelation effects minimized.

One such method is parallel tempering Monte Carlo
\cite{swendsen:86,geyer:91,marinari:92,hukushima:96,comment:name}
that has proven to be a versatile ``workhorse'' in many fields
\cite{earl:05}. Similar to replica Monte Carlo \cite{swendsen:86},
simulated tempering \cite{marinari:92}, or extended ensemble methods
\cite{lyubartsev:92}, the algorithm aims to overcome free-energy
barriers in the free energy landscape by simulating several copies
of the target system at different temperatures.  The system can thus
escape metastable states when wandering to higher temperatures and
relax to lower temperatures again in time scales several orders of
magnitude smaller than for a simple Monte Carlo simulation at one fixed
temperature.  The method has also been combined with several other
algorithms such as genetic algorithms and related optimization methods,
molecular dynamics, cluster algorithms and quantum Monte Carlo.

\subsection{Outline of the algorithm}

$M$ noninteracting copies of the system are simulated in parallel
at different temperatures $\{T_1, T_2, \ldots, T_M\}$. After a
fixed number of Monte Carlo sweeps (generally one lattice sweep)
two copies at neighboring temperatures $T_i$ and $T_{i+1}$ are
exchanged with a Monte Carlo like move and accepted with a probability
\begin{equation}
  \mathcal{T}[(E_i, T_i) \rightarrow (E_{i+1}, T_{i+1})] 
  = \min \left\{ 1,  \exp[(E_{i+1} -E_i)(1/T_{i+1} - 1/T_i)]  \right\}\;.
\label{eq:pt}
\end{equation}
A given configuration will thus perform a random walk in temperature
space overcoming free energy barriers by wandering to high
temperatures where equilibration is rapid and configurations change
more rapidly, and returning to low temperatures where relaxation times
can be long. Unlike for simple Monte Carlo, the system can efficiently
explore the complex energy landscape. Note that the update probability
in Eq.~(\ref{eq:pt}) obeys detailed balance.

At first sight it might seem wasteful to simulate a system at multiple
temperatures. In most cases, the number of temperatures does not
exceed 100 values, yet the speedup attained can be $5$ -- $6$ orders
of magnitude. Furthermore, one often needs the temperature dependence
of a given observable and so the method delivers data for different
temperatures in one run. A simple implementation of the parallel
tempering move called after a certain number of lattice sweeps using
pseudo code is shown below.

\newpage

\SourceCodeLines{9}
\begin{Verbatim}[fontsize=\small]
 algorithm parallel_tempering(*energy,*temp,*spins)

     for(i = 1 ... (num_temps - 1)) do
         delta = (1/temp[i] - 1/temp[i+1])*(energy[i] - energy[i+1])
         if(rand(0,1) < exp(delta)) then
             swap(spins[i],spins[i+1])
             swap(energy[i],energy[i+1])
         fi
     done
\end{Verbatim}

\noindent The \texttt{swap( )} function swaps neighboring energies
and spin configurations (\texttt{*spins}) if the move is accepted.
As simple as the algorithm is, some fine tuning has to be performed
for it to operate optimally.

\subsection{Selecting the temperatures}

There are {\em many} recipes on how to ideally select the position
of the temperatures for parallel tempering Monte Carlo to perform
optimally. Clearly, when the temperatures are too far apart, the energy
distributions at the individual temperatures will not overlap enough
and many moves will be rejected. The result is thus $M$ independent
simple Monte Carlo simulations run in parallel with no speed increase
of any sort. If the temperatures are too close, CPU time is wasted.

A measure for the efficiency of a system copy to traverse the
temperature space is the probability (as a function of temperature)
that a swap is accepted.  A good rule of thumb is to ensure that the
acceptance probabilities are approximately independent of temperature,
between approximately 20 -- 80\%, and do not show large fluctuations
as these would signify the breaking-up of the random walk into segments of
the temperature space. Following the aforementioned recipe, parallel
tempering Monte Carlo already outperforms any simple sampling Monte
Carlo method in a rough energy landscape. Still, the performance can
be further increased, as outlined below.

\paragraph{Traditional approaches} As mentioned before, a reasonable
performance of the algorithm can be obtained when the acceptance
probabilities are approximately independent of temperature.  If the
specific heat of a system is not strongly divergent at the phase
transition---as it is the case with spin glasses---a good starting
point is given by a {\em geometric progression} of temperatures.
Given a temperature range $[T_1, T_M]$, the intermediate $M-2$
temperatures can be computed via
\begin{equation}
T_k = T_1\prod_{i = 1}^{k - 1} R_i, \;\;\;\;\;\;\;\;\;\;\;\;
 R_i = \sqrt[M-1]{\frac{T_M}{T_1}}\;.
\label{eq:geometric}
\end{equation}
Because relaxation is slower for lower temperatures, the geometric
progression peaks the number of temperatures close to $T_1$.  If,
however, the specific heat of the system has a strong divergence, this
approach is not optimal. 

One can show that the acceptance probabilities are inversely
correlated to the functional behavior of the specific heat
per spin $c_V$ via $\Delta T_{i,i+1} \sim c_{V}T_i/\sqrt{N}$
\cite{predescu:04}. Therefore, if $c_V$ diverges, the acceptance
probabilities for a geometric temperature set show a pronounced dip at
the transition temperature. More complex methods such as the approach
by Kofke \cite{kofke:02,kofke:04}, its improvement by Rathore {\em et
al.}~\cite{rathore:05}, as well as the method suggested by Predescu
{\em et al.} \cite{predescu:04,predescu:05} aim to obtain acceptance
probabilities for the parallel tempering moves that are independent
of temperature by compensating for the effects of the specific heat.

Finally, the number of temperatures needed can be estimated via the
behavior of the specific heat. One can show that $M \sim \sqrt{N^{1
-d\nu/\alpha}}$ \cite{hukushima:96}. Here $d$ is the space dimension,
$N$ the number of spins, $\nu$ the critical exponent of the correlation
length and $\alpha$ the critical exponent of the specific heat.

In practice, it is straightforward to tune a temperature set produced
initially via a geometric progression by adding interstitial
temperatures where the acceptance rates are low.  A quick simulation
for only a few Monte Carlo sweeps yields enough information about the
acceptance probabilities to tune the temperature set by hand without
having to resort to a full equilibrium simulation.

\paragraph{Improved approaches} Recently, a new iterative feedback
method has been introduced to optimize the position of the temperatures
in parallel tempering simulations \cite{katzgraber:06a}.  The idea
is to treat the set of temperatures as an ensemble and thus
use ensemble optimization methods \cite{trebst:04} to improve
the round-trip times of a given system copy in temperature space.
Unlike the conventional approaches, resources are allocated to the
bottlenecks of the simulation, i.e., phase transitions and ground
states where relaxation is slow. As a consequence, acceptance
probabilities are temperature-dependent because more temperatures
are allocated to the bottlenecks.  The approach requires one to
gather enough round-trip data for the temperature sets to converge
and thus is not always practical. For details on the implementation,
see Refs.~\cite{katzgraber:06a} and \cite{trebst:07}, as well
as Ref.~\cite{hamze:10} for an improved version.

A similar approach to optimize the efficiency of parallel
tempering has recently been introduced by Bittner {\em et
al.}~\cite{bittner:08}. Unlike the previously-mentioned feedback
method, this approach leaves the position of the temperatures untouched
but with an average acceptance probability of 50\%.  To deal with
free-energy barriers in the simulation, the autocorrelation times of
the simulation {\em without} parallel tempering have to be measured
{\em ahead} of time. The number of MCS between parallel tempering
updates is then dependent on the autocorrelation times, i.e., close
to a phase transition, more MCS between parallel tempering moves are
performed.  Again, the method is thus optimized because resources
are reallocated to where they are needed most. Unfortunately, this
approach also requires a simulation to be done ahead of time to
estimate the autocorrelation times, but a rough estimate is sufficient.

\subsection{Example: Application to spin glasses}

To illustrate the advantages of parallel tempering over simple Monte
Carlo, we show data for a three-dimensional Ising spin glass with
Normal-distributed disorder. In that case, one can use an exact
relationship between the energy and a fourth-order spin correlator
known as the link overlap $q_\ell$ \cite{katzgraber:01}. The link
overlap is given by
\begin{equation}
q_\ell = \frac{1}{dN} \sum_{\langle i, j \rangle}
S_i^{\alpha} S_j^{\alpha} 
S_i^{\beta} S_j^{\beta}\;.
\label{eq:ql}
\end{equation}
The sum in Eq.~(\ref{eq:ql}) is over neighboring spin pairs and the
normalization is over all bonds.  If a domain of spins in a spin
glass is flipped, the link overlap measures the average length of
the boundary of the domain.

\begin{SCfigure}[][!htb]
  \includegraphics[scale=0.30]{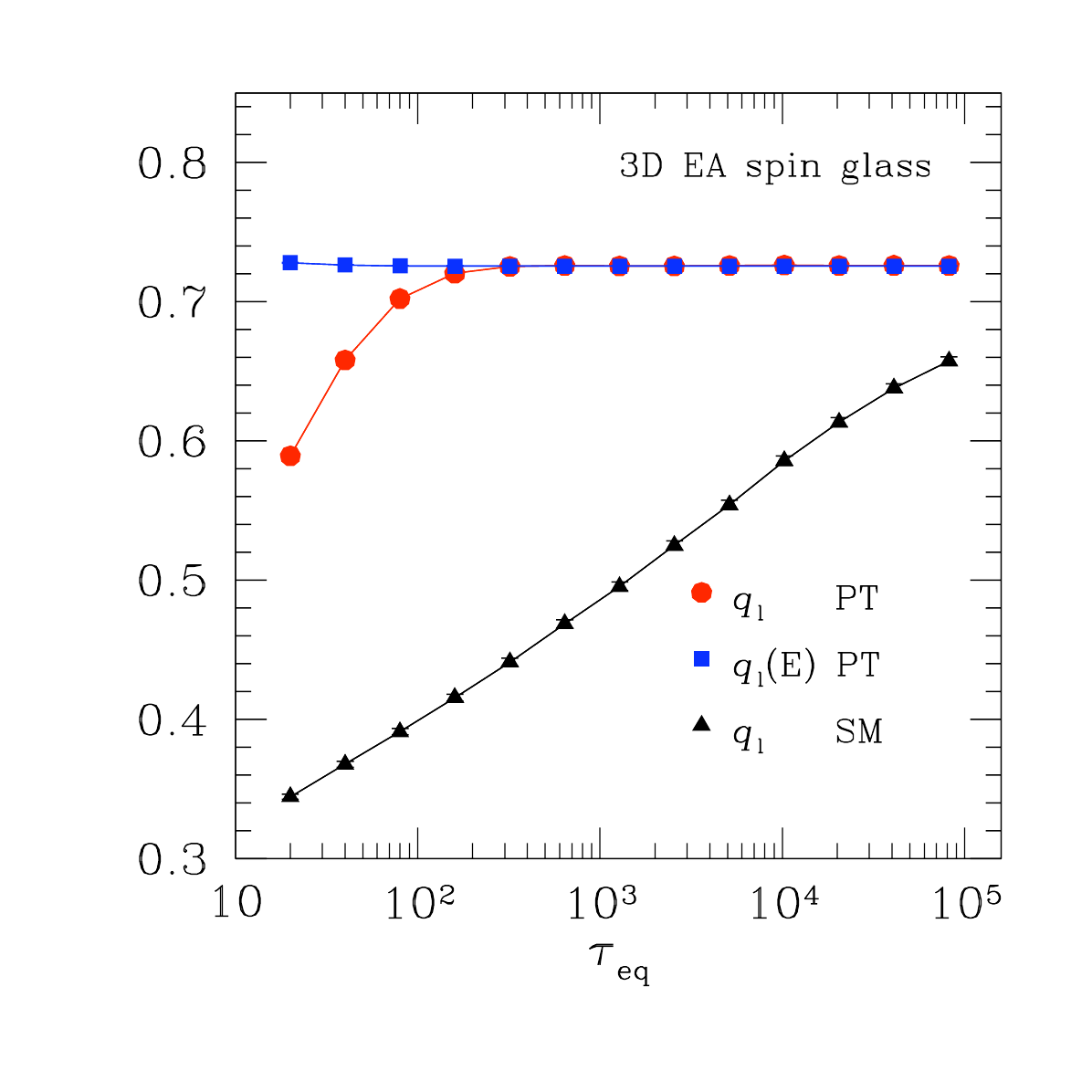}\hspace{1pc}
  \caption
   {
    Equilibration test for spin glasses with Gaussian disorder. Data
    for the link overlap (circles) have to equate to data for the link
    overlap computed from the energy (squares). This is the case after
    approximately 300 MCS when parallel tempering is used. A direct
    calculation of the link overlap using simple Monte Carlo (triangles)
    is not equilibrated after $10^5$ MCS. Data for $L = 4$, $d = 3$,
    $5000$ samples, and $T = 0.50$.
    \vspace{3pc}
   }
  \label{fig:pt}
\end{SCfigure}

The internal energy per spin $u$ is given by
\begin{equation}
u = - \frac{1}{N} \sum_{\langle i,j \rangle} [\, J_{ij} \langle S_i S_j
\rangle]_{\rm av}\;,
\end{equation}
where $\langle \cdots \rangle$ represents the Monte Carlo average for
a given set of bonds, and $[ \cdots ]_{\rm av}$ denotes an average over the
(Gaussian) disorder.  One can perform an integration by parts over
$J_{ij}$ to relate $u$ to the average link overlap defined in
Eq.~(\ref{eq:ql}), i.e.,
\begin{equation}
[\langle q_\ell \rangle]_{\rm av} = 1 + \frac{Tu}{d}\;.
\label{eq:equilc}
\end{equation}
The simulation starts with a random spin configuration.  This means
that the two sides of Eq.~(\ref{eq:equilc}) approach equilibrium
from {\em opposite} directions. Data for $q_\ell$ will be too
small because we started from a random configuration, whereas the
initial energy will not be as negative as in thermal equilibrium.
Once both sides of Eq.~(\ref{eq:equilc}) agree, the system is in
thermal equilibrium.  This is illustrated in Fig.~\ref{fig:pt} for
the Edwards-Anderson Ising spin glass with $4^3$ spins and $T = 0.5$
which is approximately 50\% $T_c$. The data are averaged over 5000
realizations of the disorder. While the data for $q_\ell$ generated with
parallel tempering Monte Carlo agree after approximately 300 MCS, the
data produced with simple Monte Carlo have not reached equilibrium even
after $10^5$ MCS, thus illustrating the power of parallel tempering
for systems with a rough energy landscape.

\section{Rare events: Probing tails of energy distributions}

When computing distribution functions (histograms) $P(x)$ of
a quantity $x$ typically simple-sampling techniques are used
\cite{comment:dis}. The quantity $x$ can be an order parameter,
a free energy, an internal energy, a matching probability, etc. In
these simple-sampling techniques $N_{\rm samp}$ instances are computed
and subsequently binned in order to obtain the desired distribution.
If $N_{\rm samp}$ samples are computed, then the maximal ``resolution''
of a bin is $\sim 1/N_{\rm samp}$, and thus, for example, $\sim 10^7$
samples have to be computed to resolve seven orders of magnitude in
a histogram. If, however, the tails need to be probed to 18 order is
magnitude precision, then the computations become quickly intractable
because $N_{\rm samp} \sim 10^{18}$ samples have to be computed.
One alternative is to use multicanonical methods \cite{berg:91,berg:92}
that, for example, have been used before to overcome the limitations
of simple-sampling techniques in order to probe tails of overlap
distribution functions in spin glasses \cite{berg:02,berg:02a}.

Here we outline a method related to multicanonical approaches
based on ideas presented in Ref.~\cite{hartmann:02e} that also
overcomes the limitations of simple-sampling techniques and works
for systems with disorder, e.g., spin glasses. The idea is to
perform an importance-sampling simulation of $P(x)$ \textit{in the
disorder} with a {\em guiding function} estimated from simple-sampling
simulations. Similar approaches have been used before in the studies
of distributions of sequence alignment scores~\cite{hartmann:02e},
free-energy barriers in the Sherrington-Kirkpatrick model
\cite{bittner:06}, as well as fluctuations in classical
magnets~\cite{hilfer:03} (albeit the latter without disorder).

\subsection{Case study: Ground-state energy distributions}

A disordered system is defined by a Hamiltonian
$\mathcal{H}_\mathcal{J}(\mathcal{C})$, where the disorder
configuration $\mathcal{J}$ is chosen from a probability distribution
${\mathcal P}(\mathcal{J})$ and $\mathcal{C}$ denotes the phase-space
configuration of the system. The ground-state energy $E$ of a given
disorder configuration $\mathcal{J}$ is defined by
\begin{equation}
E(\mathcal{J}) = \min_{\mathcal{C}} \, \mathcal{H}_\mathcal{J}(\mathcal{C}).
\label{eminop}
\end{equation}
Together with the disorder distribution ${\mathcal P}(\mathcal{J})$,
this defines the ground-state energy distribution
\begin{equation}
P(E) = \int d{\mathcal J} \, {\mathcal P}(\mathcal{J}) \, 
\delta \left[ E - E(\mathcal{J}) \right].
\end{equation}

\subsection{Simple sampling}

$N_\mathrm{samp}$ independent disorder configurations $\mathcal{J}_i$
are chosen from ${\mathcal P}(\mathcal{J})$ and the ground-state energy
is calculated for each disorder configuration. The calculation of
the ground-state energy in itself is a difficult optimization problem
that we sweep under the rug (see Refs.~\cite{hartmann:01,hartmann:04}
for efficient methods).  From the ground-state energies of these
disorder configurations, the ground-state energy distribution can be
estimated via
\begin{equation}
P(E) = \frac{1}{N_\mathrm{samp}} \sum_{i=1}^{N_\mathrm{samp}} \delta 
\left[ E - E(\mathcal{J}_i) \right],
\label{pexp}
\end{equation}
so that the averages of functions with respect to the disorder are
replaced by averages with respect to the $N_\mathrm{samp}$ random
samples. The functional form of the ground-state energy distribution
and its parameters can be estimated by a maximum likelihood fit of
an empirical distribution $F_\theta(E)$ with parameters $\{\theta\}$
to the data~\cite{cowan:98}. Note that due to the limited range
of energies sampled by the simple-sampling algorithm it is often
difficult or even impossible to quantify how well the tails of the
distribution are described by a maximum-likelihood fit.

\subsection{Importance sampling with a guiding function}

Assume it is easy to find a function $F_{\theta}(E)$ that accurately
describes the ground-state energy distribution calculated from a quick
simple-sampling simulation as described in the previous section. In
that case, an importance-sampling Monte Carlo algorithm \textit{in
the disorder}~\cite{newman:99,landau:00,hartmann:02e} can be used
to probe the tails efficiently by using $F_{\theta}(E)$ as a guiding
function.  We start from a random disorder configuration $\mathcal{J}
= \mathcal{J}_0$ with ground-state energy $E(\mathcal{J}_0)$. From
the $i$-th configuration $\mathcal{J}_i$, we generate the $i+1$-th
configuration $\mathcal{J}_{i+1}$ via a Metropolis-type update:
\begin{enumerate}

\item Select a disorder configuration $\mathcal{J}'$ by replacing
a subset of $\mathcal{J}$ chosen at random (e.g., a single bond
chosen at random) with values chosen according to $P(\mathcal{J})$
and calculate its ground-state energy $E(\mathcal{J}')$.

\item Set $\mathcal{J}_{i+1} = \mathcal{J}'$ with probability
\begin{equation}
P_\mathrm{accept} = \min \left\{ 
\frac{F_{\theta}
\left[E(\mathcal{J}_i)\right]}{F_{\theta}\left[E(\mathcal{J'})\right]}
, 1 \right\}
\label{paccept}
\end{equation}
and $\mathcal{J}_{i+1} = \mathcal{J}_i$ otherwise.

\end{enumerate}
With this algorithm a disorder configuration $\mathcal{J}$ is
visited with probability $1/F_{\theta}[E(\mathcal{J})]$, such that
the probability to visit a disorder configuration with ground-state
energy $E$ is $P(E)/F_{\theta}(E)$. If $F_{\theta}(E) \sim P(E)$,
then each energy is visited with the {\em same} probability resulting
in a flat-histogram sampling of the ground-state energy distribution.
To prevent trapping of the algorithm in an extremal region of phase
space the range of energies that the algorithm is allowed to visit
can be restricted (see Ref.~\cite{koerner:06} for details).

Note that successive configurations visited by the algorithm are not
independent. To ensure that the data are not correlated, only samples
each $\tau$ measurements are considered in the average, where $\tau$
is the exponential autocorrelation time of the energy. It can be
computed from the energy autocorrelation function
\begin{equation}
\zeta_{\rm auto}(i) = \frac{\langle E_j^{} \, E_{j+i}^{} \rangle - 
\langle E_j^{} \rangle \, \langle E_{j+i}^{} \rangle}{\langle E_j^2
\rangle - 
\langle E_j^{} \rangle_{}^2},
\label{actime}
\end{equation}
where it decays to $1/e$ \cite{landau:00}. Here $E_i$ is the ground
state energy after the $i$-th Monte Carlo step and $\langle \ldots
\rangle$ represents an average over Monte Carlo time.  To be sure
that the visited ground-state configurations are not correlated,
we empirically only use every $4\tau$-th measurement.  Once the
autocorrelation effects have been quantified, the data can be
analyzed with the same methods as the simple-sampling results [see
Eq.~(\ref{pexp})].

\begin{figure}[!htb]
\centering
\includegraphics[scale=0.50]{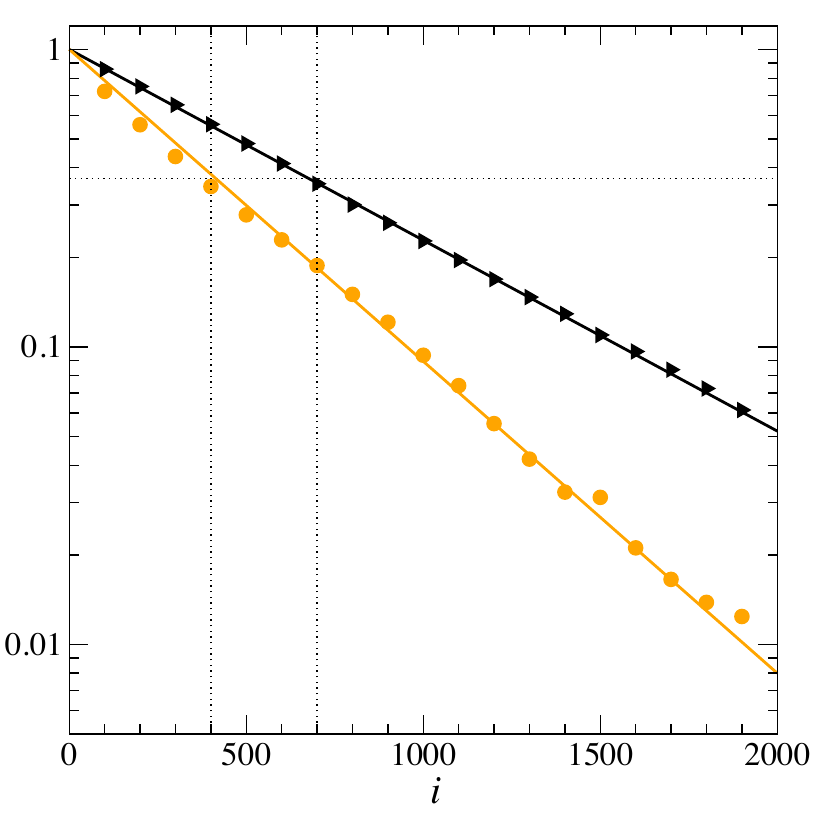}
\caption
{
Autocorrelation function as defined in Eq.~(\ref{actime}) for the
Sherrington-Kirkpatrick spin glass and system sizes $N=16$ (circles)
and $N=128$ (triangles).  The value $1/e$ is marked by the horizontal
dotted line. Time steps $i$ are measured in Monte Carlo steps. (Figure
adapted from Reference \cite{koerner:06}).
}
\label{fig:auto}
\end{figure}

\subsection{Example: The Sherrington-Kirkpatrick Ising spin glass}

The Sherrington-Kirkpatrick \cite{sherrington:75} model is given by
the Hamiltonian
\begin{equation}
\mathcal{H}_\mathcal{J}(\{S_i\}) = \sum_{ i < j } J_{ij} \, S_i \, S_j,
\label{skham}
\end{equation}
where the $S_i = \pm 1$ ($i=1,\ldots,N$) are Ising spins, and
the interactions $\mathcal{J} = \{J_{ij}\}$ are identically and
independently distributed random variables chosen from a Normal
distribution with zero mean and standard deviation $(N-1)^{-1/2}$. The
sum is over all spins in the system, i.e., the model represents
the mean-field version of the Edwards-Anderson Ising spin glass
introduced before.

For the SK model several optimization algorithms, such es
extremal optimization \cite{boettcher:01}, hysteretic optimization
\cite{pal:06}, as well as other algorithms such as genetic and Bayesian
algorithms \cite{hartmann:01,hartmann:04}, and even parallel tempering
\cite{hukushima:96} can be used to estimate ground-state energies
for small to moderate system sizes.

We first compute $10^5$ ground-state energies and bin the data into 50
bins and perform a maximum-likelihood fit to a function that describes
the shape of the ground-state energy distribution best. In this
case, this is a modified Gumbel distribution \cite{gumbel:60}:
\begin{equation}
F_{\mu,\nu,m}(E) \propto \exp \left[ 
m \, \frac{E-\mu}{\nu}  - m \, \exp \left( \frac{E-\mu}{\nu} \right)
\right].
\label{modgumbel}
\end{equation}
The modified Gumbel distribution is parametrized by the ``location''
parameter $\mu$, the ``width'' parameter $\nu$, and the ``slope''
parameter $m$.  The parameters $\mu$, $\nu$ and $m$ estimated from a
maximum-likelihood fit represent the input parameters for the guiding
function used in the importance-sampling simulation in the disorder.
To perform a step in the Monte Carlo algorithm, we choose a site at
random, replace all bonds connected to this site (the expected change
in the ground-state energy is then of the order $\sim 1/N$), calculate
the ground-state energy of the new configuration, and accept the new
configuration with the probability given in Eq.~(\ref{paccept}).
A study of the energy autocorrelation shows that for system sizes
between 16 and 128 spins the autocorrelation times are of the order
of 400 to 700 Monte Carlo steps, see Fig.~\ref{fig:auto}.

\begin{figure}[!htb]
\centering
\includegraphics[scale=0.50]{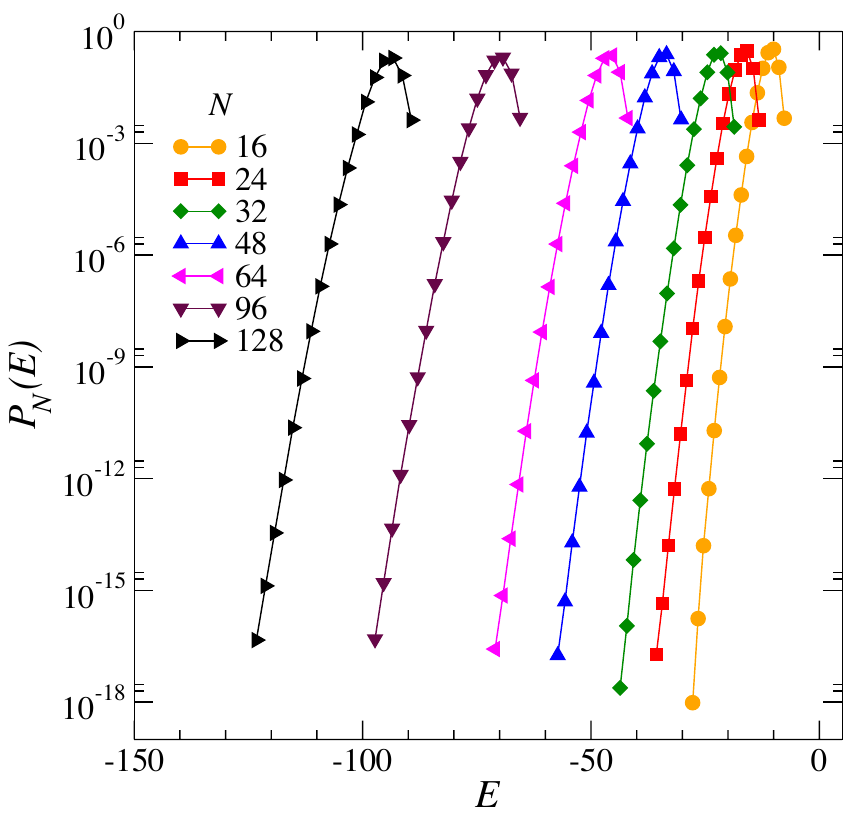}
\caption
{
Ground-state energy distributions of the Sherrington-Kirkpatrick
model for different system sizes obtained from a importance-sampling
simulation with a guiding function.  Although only $\sim 10^3$
samples per system size $N$ were simulated, the resolution of the
histograms is up to 18 orders of magnitude. (Figure adapted from
Reference \cite{koerner:06}).
}
\label{fig:distr}
\end{figure}

Figure \ref{fig:distr} shows the energy distributions for the
Sherrington-Kirkpatrick Ising spin glass for different system sizes
$N$. The data span up to 18 orders of magnitude and were produced with
approximately 1000 samples per system size, therefore illustrating
the immense power of importance sampling simulations when probing
tails of distributions.  This would be {\em impossible} to obtain
with simple-sampling techniques.

In comparison to similar methods \cite{hartmann:02e,hilfer:03} the
presented approach has several advantages due to its simplicity:
Instead of iterating towards a good guiding function, which may
be quite expensive computationally, we use a maximum likelihood
fit as a guiding function. Therefore, the proposed algorithm is
straightforward to implement and considerably more efficient than
traditional approaches, provided a good guiding function, i.e.,
a good maximum-likelihood fit to the simple-sampling results, can be
found. Note also that the method can be generalized to any distribution
function, such as an order-parameter distribution.

\section{Other Monte Carlo methods}

In addition to the Monte Carlo methods outlined, there is a
vast selection of other approaches based on random Monte Carlo
sampling to study physical systems.  In this section some selected
finite-temperature Monte Carlo methods are briefly outlined. The
reader is referred to the literature for details. Note that most
algorithms can be combined for better performance. For example, one
could combine parallel tempering with a cluster algorithm to speed
up simulations both around and far below the transition.

\paragraph{Cluster algorithms} In addition to the Wolff cluster
algorithm \cite{wolff:89} outlined in Sec.~\ref{subsec:wolff},
the Swendsen-Wang \cite{swendsen:87} algorithm also greatly helps
to overcome critical slowing down of simulations close to phase
transitions. There are also specially-crafted cluster algorithms for
spin glasses \cite{houdayer:01}.

\paragraph{Simulated annealing} Simulated annealing is probably
the simplest heuristic \linebreak ground-state search approach. A
Monte Carlo simulation is performed until the system is in thermal
equilibrium. Subsequently, the temperature is quenched according to
a pre-defined protocol until $T$ close to zero is reached. After each
quench, the system is equilibrated with simple Monte Carlo. The system
should converge to the ground state, although there is no guarantee
that the system will not be stuck in a metastable state.

\paragraph{Flat-histogram methods} Flat-histogram algorithms include
the multicanonical method \cite{berg:91,berg:92}, broad histograms
\cite{deoliveira:96} and transition matrix Monte Carlo \cite{wang:02}
when combined with entropic sampling, as well as the adaptive algorithm
of Wang and Landau \cite{wang:01,wang:01a} and its improvement
by Trebst {\em et al.}~\cite{trebst:04}.  The advantage of these
algorithms is that they allow for an estimate of the free energy;
this is usually not available from standard Monte Carlo methods.

\paragraph{Quantum Monte Carlo} In addition to the aforementioned
Monte Carlo methods that treat classical problems, quantum extensions
such as variational Monte Carlo, path integral Monte Carlo, etc.~have
been developed for quantum systems \cite{suzuki:93}.
\bigskip

\section*{Acknowledgments}

I would like to thank Juan Carlos Andresen and Ruben Andrist for
critically reading the manuscript. Furthermore, I thank M.~Hasenbusch
for spotting an error in Sec.~\ref{sec:traditional}.

\bibliographystyle{plain}  
\bibliography{comments,refs}

\begin{thebibliography}{10}
\bibitem{comment:sources}
The bulk of this chapter is based on material collected from the different
  books cited.
\bibitem{comment:alk}
The alcohol is to improve the randomness of the sampling. If the
  experimentalist is not of legal drinking age, it is recommended to close the
  eyes and rotate 42 times on the spot at high speed before a pebble is thrown.
\bibitem{comment:pseudo}
The pseudo code used does not follow any rules and is by no means consistent.
  But it should bring the general ideas across.
\bibitem{comment:mc}
Although the algorithm is known as the {\em Metropolis algorithm},
  N.~Metropolis' contribution to the project was minimal. He merely was the
  team leader at the lab. The bulk of the work was carried out by two couples,
  the Rosenbluths and the Tellers.
\bibitem{comment:dis}
This section is based on work published in Ref.~\cite{koerner:06}.
\end{thebibliography}

\end{document}